\documentclass[twocolumn,showpacs,preprintnumbers,amsmath,amssymb]{revtex4}

\usepackage{dcolumn}
\usepackage{bm}

\usepackage{hyperref}

\usepackage[dvips]{graphicx}

\newcommand{\eh}{\ensuremath{158 A}~GeV }

\newcommand{\enI}{\ensuremath{20 A}~GeV }

\newcommand{\Pb}{Pb+Pb~}

\begin{document}

\preprint{APS/123-QED}

\title{Searching for the Critical Point of QCD:\\ Theoretical Benchmark Calculations}

\author{Benjamin Lungwitz}
\email{lungwitz@ifk.uni-frankfurt.de}
\affiliation{%
Institut f\"ur Kernphysik, Johann Wolfgang Goethe Universit\"at, Max-von-Laue-Str.~1, D-60438 Frankfurt am Main, Germany
}%

\author{Marcus Bleicher}%
\email{bleicher@th.physik.uni-frankfurt.de}
\affiliation{%
Institut f\"ur Theoretische Physik, Johann Wolfgang Goethe-Universit\"at, Max-von-Laue-Str.~1, D-60438 Frankfurt
 am Main, Germany
}%

\date{\today}

\begin{abstract}
We present a comprehensive study of event-by-event multiplicity fluctuations in nucleon-nucleon and 
nucleus-nucleus interactions from AGS/FAIR to RHIC energies within the UrQMD transport approach. 
The scaled variances of negative, positive, and all charged hadrons are analysed. 
The scaled variance in central \Pb collisions increases with energy and behaves similar to inelastic p+p interactions.
We find a non-trivial dependence of multiplicity fluctuations on the rapidity and transverse momentum interval
used for the analysis and on the centrality selection procedure.
Quantitative predictions for the NA49 experiment are given, taking into account the acceptance
of the detector and the selection procedure of central events.
\end{abstract}

\pacs{25.75.Nq,24.60.-k,12.38.Mh}

\keywords{Event-by-event, multiplicity fluctuations, scaled variance}

\maketitle

\section{Introduction}

At high energy densities ($\approx 1$ $\rm{GeV/fm^3}$) a phase transition from a hadron gas to a quark-gluon-plasma (QGP) 
is expected to occur. There are indications that at RHIC and top SPS energies a quark-gluon-plasma is created 
in the early stages of heavy ion collisions~\cite{Heinz:2000bk, Gyulassy:2004zy}.
And indeed, the energy dependence of various observables show anomalies at low SPS energies which might be 
related to the onset of deconfinement~\cite{Gazdzicki:2004ef,{Gazdzicki:1998vd}}. 

While several observables \cite{Bass:1998vz} have been proposed throughout the last decades
to study the characteristics of the highly excited matter
created in heavy ions collisions the ones related to fluctuations and correlations
seem to be the most prospective. Fluctuation probes might be more adequate for the
exploration of heavy ion reactions, because the distributions of energy density
or initial temperature, isospin and particle density have strong fluctuations
from event to event \cite{Stodolsky:1995ds,Shuryak:1997yj,Bleicher:1998wd}.
On the theoretical side event-by-event fluctuations were suggested to study
\begin{itemize}
\item
kinetic and chemical equilibration in nuclear collisions \cite{Gazdzicki:1992ri,Mrowczynski:1997kz,Bleicher:1998wu,Mrowczynski:1999sf,Capella:1999uc,Mrowczynski:1999un,Mrowczynski:2004cg,Sa:2001ma,Mekjian:2004qf},
\item
the onset of the deconfinement phase \cite{Asakawa:2000wh,Muller:2001wj,Bleicher:2000ek,
Koch:2001zn,Jeon:2003gk,Shi:2005rc,Jeon:2005kj,Gazdzicki:1998vd,Gazdzicki:2003bb}
\item
the location of the tri-critical end-point of the QCD phase transition \cite{Stephanov:1998dy,Stephanov:1999zu,Hatta:2003wn} or
\item
the formation of exotic states, like DCCs \cite{Bleicher:2000tr}.
\end{itemize}
On the experimental side, progress has been made by many experiments
to extract momentum and particle number ratio fluctuations from
heavy ion reaction:  Currently, event-by-event fluctuations are actively studied in the SPS energy regime 
(starting from 20A~GeV on) by the NA49 group \cite{Appelshauser:1999ft,Reid:1999it,Afanasev:2000fu,
Anticic:2003fd,Roland:2004pu,Alt:2004ir,Alt:2004gx,Alt:2006jr,Roland:2005pr}, the 
CERES \cite{Adamova:2003pz,Sako:2004pw,Appelshauser:2004xj,Appelshauser:2004ms} and the WA98 collaboration \cite{Aggarwal:2001aa}.
At RHIC energies the PHENIX \cite{Adcox:2002mm,Adcox:2002pa,Adler:2003xq} and
STAR \cite{Adams:2003st,Adams:2003uw,Adams:2004kr} experiments
are addressing the field of single event physics.

In \cite{Gazdzicki:2003bb} it was predicted that the onset of deconfinement should lead to a non-monotonous 
behaviour in multiplicity fluctuations ("shark fin").
Also droplet formation during the phase transition is expected to produce non-statistical fluctuations 
10-100 times the Poisson expectation~\cite{Mishustin:2006ka}. 
Furthermore, lattice QCD calculations suggest the existence of a critical point in the phase diagram of 
strongly interacting matter which separates the first order phase transition from a crossover. 
Thus, if the system passes the vicinity of the critical region during its evolution and remains there for long 
enough time one expects an increase of multiplicity fluctuations \cite{Stephanov:1999zu}.

The NA49 collaboration is currently searching for such anomalies in the energy dependence of multiplicity 
fluctuations in \Pb collisions. A similar program to search for the critical point and signals for the onset of
deconfinement will be undertaken by the RHIC experiments (the planned critRHIC program ) with a 
lowering of the RHIC's beam energy towards the SPS energy regime and
the NA61 (SHINE) experiment~\cite{Antoniou:2006mh} with the focus on light ion collisions.
For the present investigation however, we will focus on the soon available data from the NA49 experiment on
multiplicity fluctuations.
Unfortunately both the geometrical acceptance of the detector and the centrality selection in the NA49 experiment 
is not trivial and have an influence on the multiplicity fluctuations. In order to observe an increase of 
fluctuations caused 
by one of the effects mentioned above, a systematic theoretical investigation within a transport approach
is needed. Only with this baseline for the expected multiplicity fluctuations within the experimental acceptance
a possible excess of fluctuations in data could be unambiguously interpreted as a signal for the critical point 
or the onset of deconfinement. The model predictions presented in this paper
are obtained using UrQMD 1.3 \cite{Bass:1998ca,Bleicher:1999xi}. 
For a complementary transport theoretical study of multiplicity fluctuations, the reader is referred to 
\cite{Konchakovski:2007mg,Konchakovski:2007ss,Konchakovski:2006aq,Konchakovski:2005hq}.

\section{Measure of Multiplicity Fluctuations}\label{c_measure}

The probability to have in an event a given number of particles $n$ in the acceptance is denoted 
as $P(n)$, with the normalisation $\sum\limits_n P(n) = 1$.

The measure of multiplicity fluctuations used in this paper is the scaled variance $\omega$ defined as
\begin{equation}
\omega=\frac{Var(n)}{\langle n\rangle}=\frac{\langle n^2\rangle-\langle n\rangle^2}{\langle n\rangle}
\end{equation}
where $Var(n)=\sum\limits_n (n-\langle n\rangle)^2 P(n)$ and $\langle n\rangle=\sum\limits_n n \cdot P(n)$ are 
the variance and the mean of the multiplicity distributions, respectively.

This measure is used because of its two properties.
Firstly in a grand-canonical statistical model neglecting quantum effects the multiplicity is a Poisson distribution:
\begin{equation}
P(n)=\frac{\langle n\rangle^n}{n!} \cdot \exp(-\langle n\rangle)
\end{equation}
The variance of a Poisson distribution is equal to its mean, the scaled variance is therefore $\omega=1$, 
independent of mean multiplicity.

Secondly in a Wounded Nucleon Model~\cite{Bialas:1976ed}, the scaled variance in $A+A$ collisions is the same 
as in proton-proton collisions provided the number of wounded nucleons is fixed.
If the particles are produced independently in momentum space, the scaled variance in a limited acceptance 
is related to the scaled variance in full phase space ($4\pi$): 
\begin{equation}\label{wscale}
\omega_{acc}=p \cdot (\omega_{4\pi} -1) +1
\end{equation}
where $p$ is the fraction of tracks which are in the corresponding acceptance.
For a small acceptance $p$ the scaled variance approaches 1.
Note that effects like resonance decays, quantum statistics and energy-momentum conservation 
introduce correlations in momentum space and 
therefore a scaling according to equation \ref{wscale} is generally not valid.

In the following the scaled variance of positively, negatively and all charged hadrons are denoted 
as $\omega(h^+)$, $\omega(h^-)$ and $\omega(h^\pm)$, respectively.

\section{The UrQMD Model}

For our investigation, we apply the Ultra-relativistic Quantum Molecular
Dynamics model (UrQMD v1.3)~\cite{Bleicher:1999xi,Bass:1998ca}
to heavy ion reactions from $E_{\rm beam}= 20A$~GeV to $E_{\rm beam}= 158A$~GeV.
This microscopic transport approach is based on the covariant propagation of
constituent quarks and di-quarks accompanied by mesonic and baryonic
degrees of freedom. It simulates multiple interactions of
in-going and newly produced particles, the excitation
and fragmentation of colour strings and the formation and decay of
hadronic resonances.
Towards higher energies, the treatment of sub-hadronic degrees of freedom is
of major importance.
In the present model, these degrees of freedom enter via
the introduction of a formation time for hadrons produced in the
fragmentation of strings \cite{Andersson:1986gw,Nilsson-Almqvist:1986rx,Sjostrand:1993yb}.
A phase transition to a quark-gluon state is
not incorporated explicitly into the model dynamics. However,
a detailed analysis of the model in equilibrium, yields an effective equation of state of
Hagedorn type \cite{Belkacem:1998gy,Bravina:1999dh}.

This model has been used before to study event-by-event fluctuations rather successfully \cite{Bleicher:1998wd,Bleicher:1998wu,Bleicher:2000ek,Bleicher:2000tr,Jeon:2005kj,Haussler:2005ei,Konchakovski:2005hq,Haussler:2006rg}  and yields a 
reasonable description of inclusive particle distributions.
For a complete review of the model, the reader is referred to \cite{Bass:1998ca,Bleicher:1999xi}.

\section{Energy Dependence of Multiplicity Fluctuations}\label{ed_s}

\begin{figure}
\includegraphics[height=5.9cm]{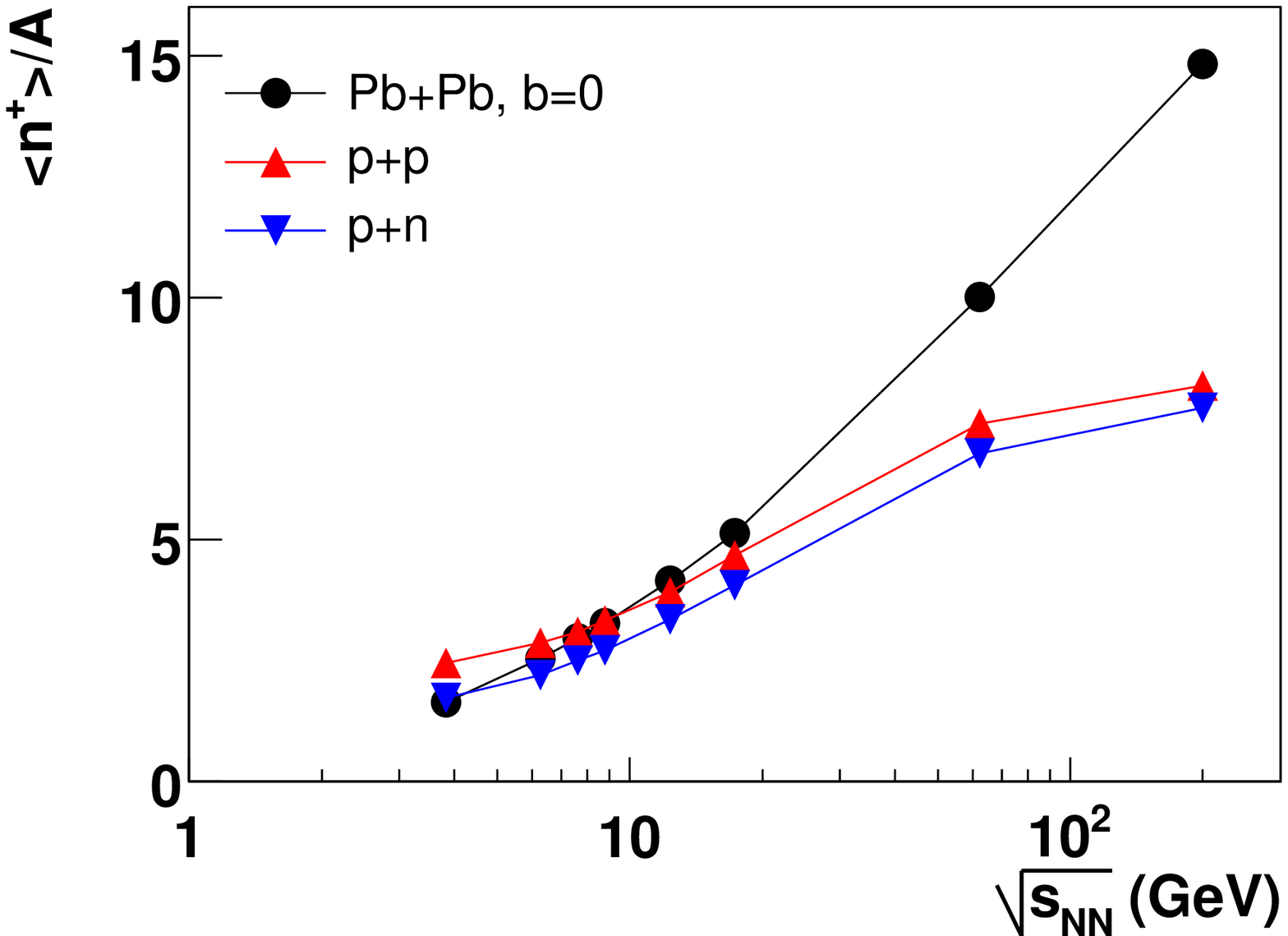}
\includegraphics[height=5.9cm]{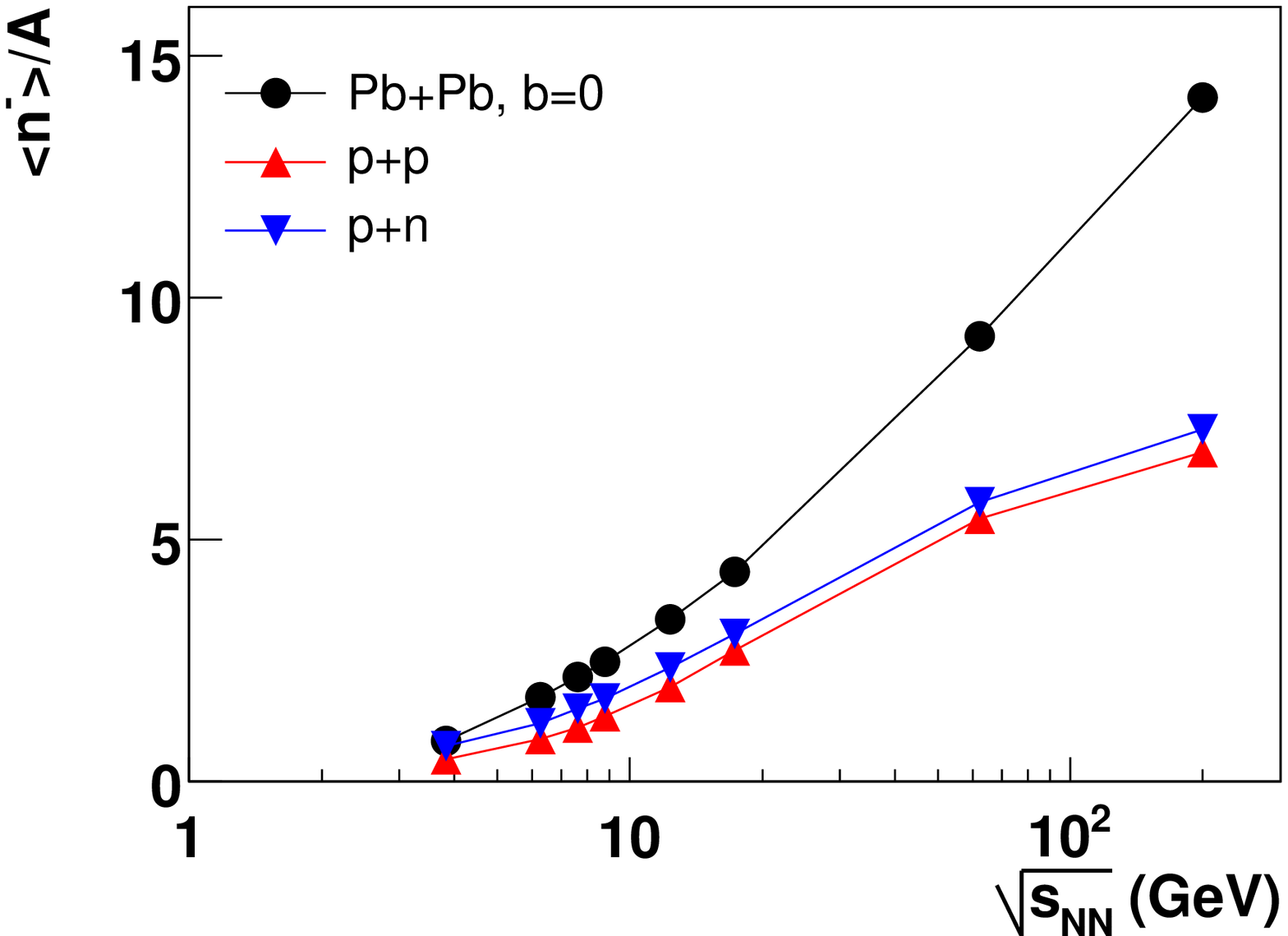}\\
\includegraphics[height=5.9cm]{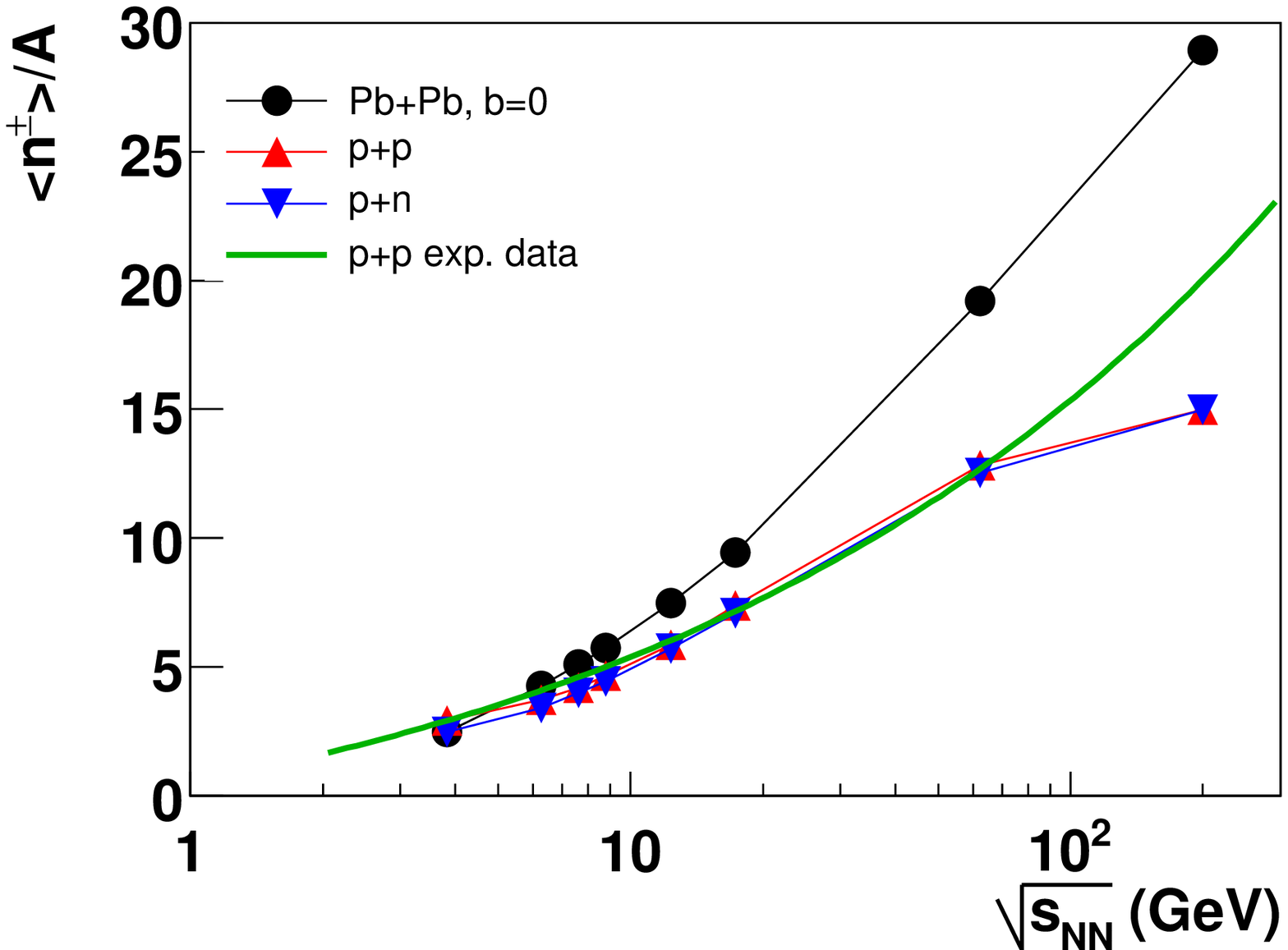}\\
\caption{\label{ed_n}(Color online) Mean multiplicity in $4\pi$ of inelastic p+p, p+n and central \Pb collisions as a function of collision energy. 
Top left: positively, top right: negatively, bottom: all charged hadrons.}
\end{figure}

The energy dependence of the mean multiplicity, normalised by the number of nucleons of one projectile ($A=1$ for p+p, p+n, $A=208$ for Pb+Pb)
of positively, negatively and all charged particles in p+p, p+n and \Pb collisions is shown in figure~\ref{ed_n}.
For p+p and p+n interactions all inelastic collisions are selected. For \Pb the impact parameter of the collisions are set to $b=0$.
The calculations were performed for AGS ($E_{lab}=6.87A$ GeV), SPS ($E_{lab}=20A$, $30A$, $40A$, $80A$ and $158A$ GeV) and RHIC ($\sqrt{s_{NN}}=62.4$ and
$200$ GeV) energies. 
In the UrQMD 1.3 model the mean multiplicity per number of projectile nucleons is significantly larger in \Pb collisions in comparison to p+n interactions.

The mean multiplicity of all charged hadrons in p+p interactions obtained by various experiments is parameterised in \cite{Heiselberg:2000fk} as
\begin{equation}
<n^\pm> \approx -4.2 + 4.69 \cdot \left( \sqrt{s_{NN}}/GeV \right)^{0.31}
\end{equation}
Except for top RHIC energies the parametrisation of the experimental data is in agreement with the UrQMD result.

\begin{figure}
\includegraphics[height=5.9cm]{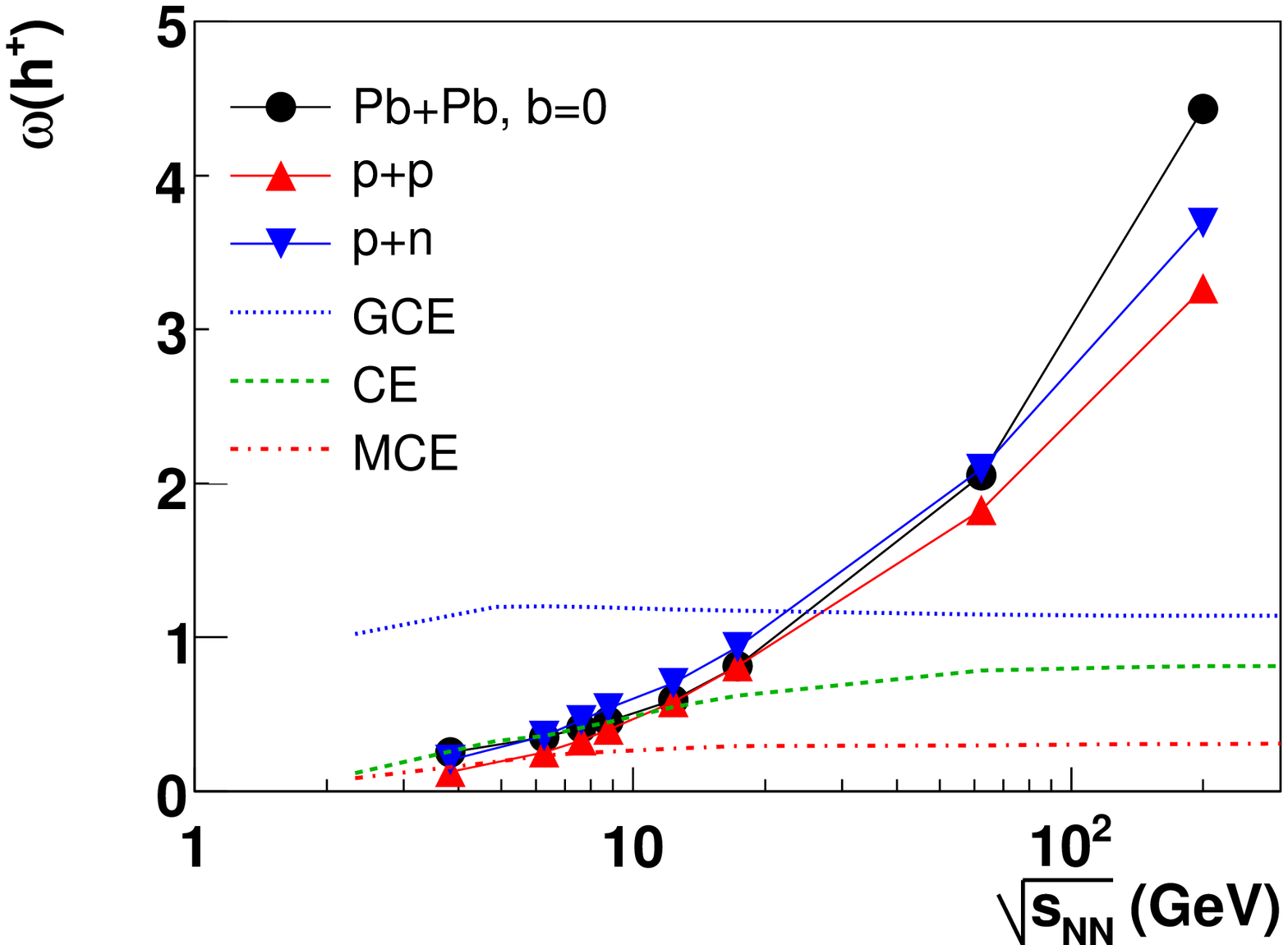}
\includegraphics[height=5.9cm]{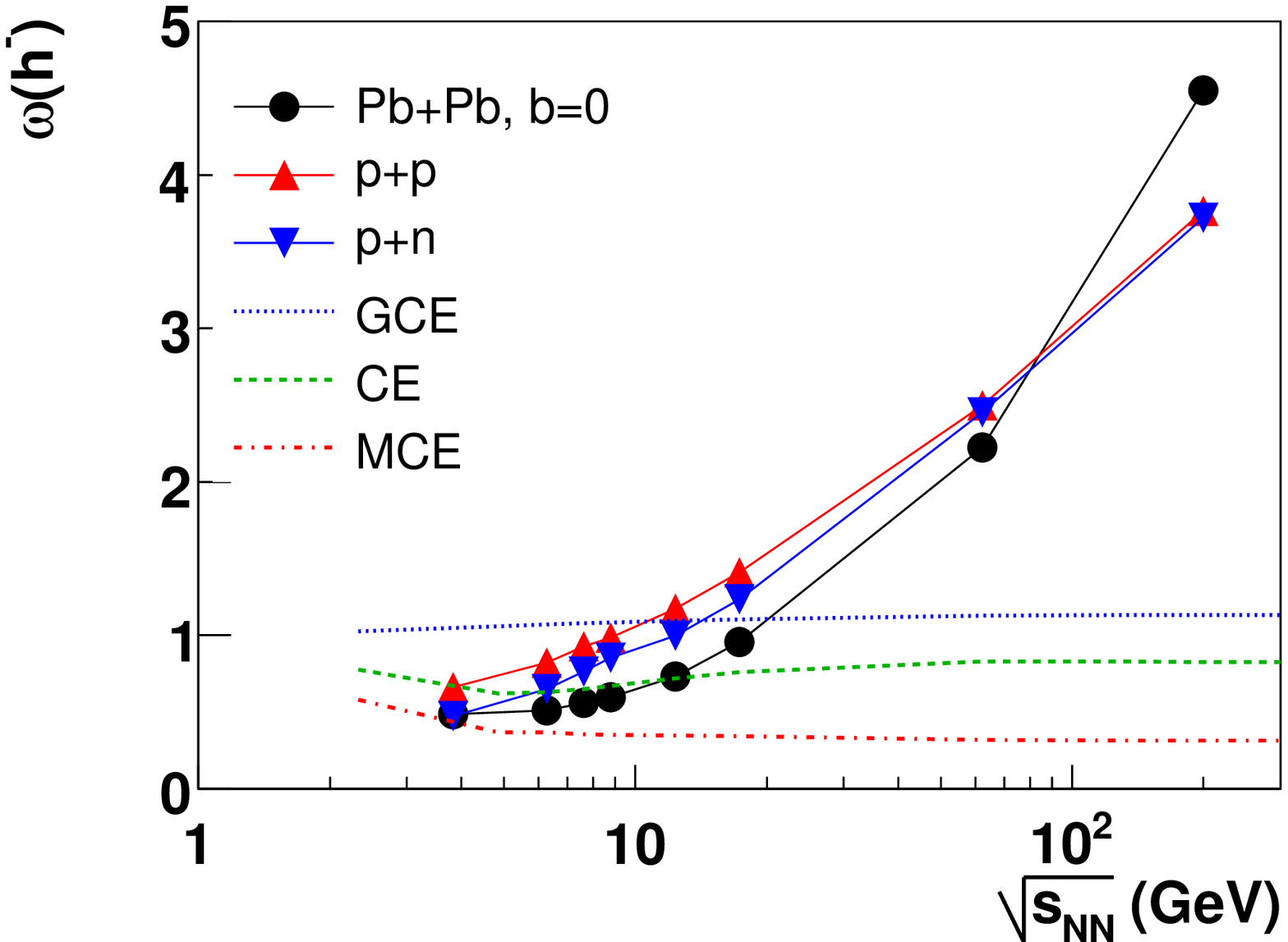}\\
\includegraphics[height=5.9cm]{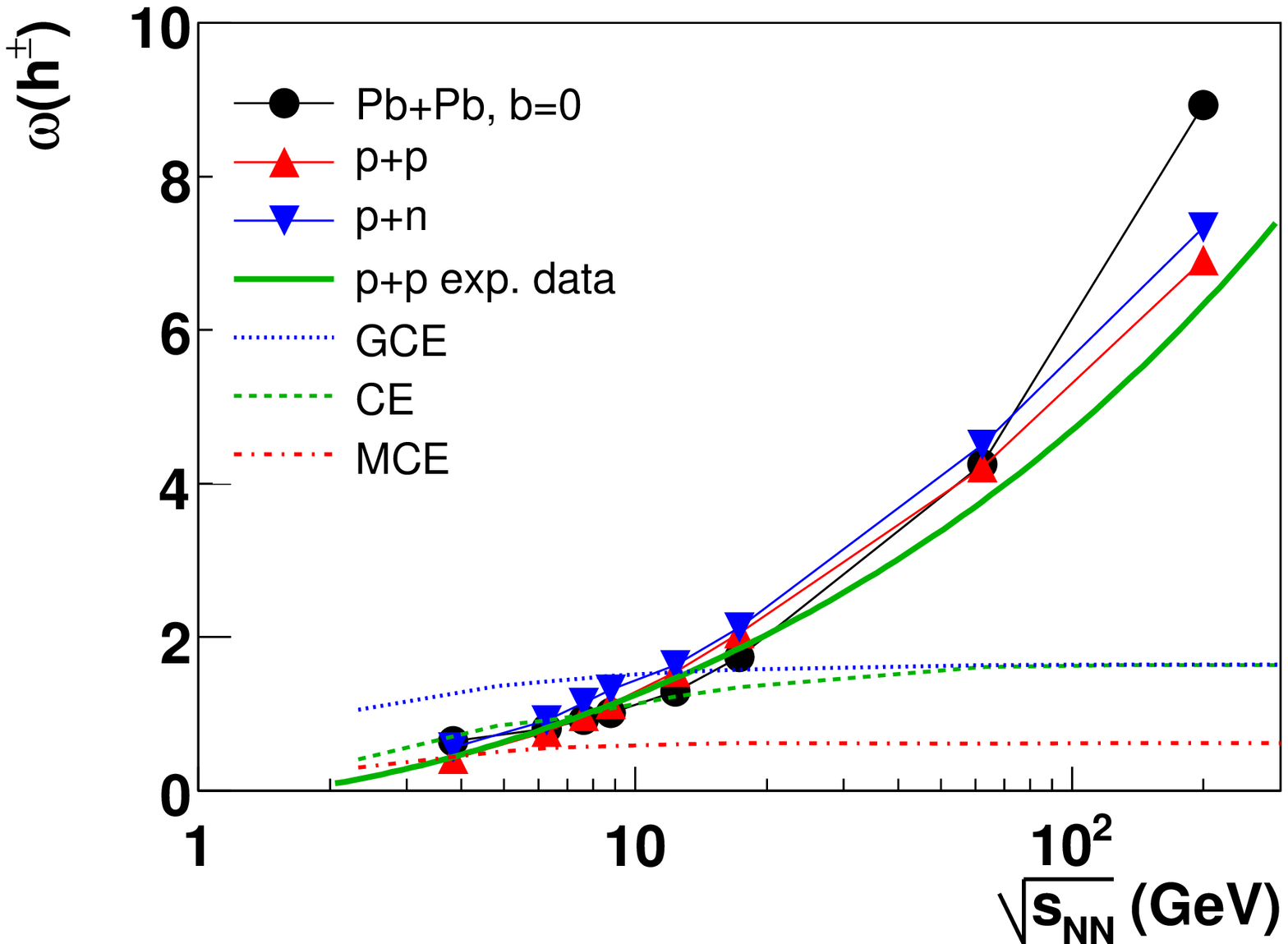}\\
\caption{\label{ed_w}(Color online) Scaled variance in $4\pi$ of inelastic p+p, p+n and central \Pb collisions as a function of collision energy in comparison to
hadron gas model predictions \cite{Begun:2006uu} 
for \Pb collisions.
Top: positively, middle: negatively, bottom: all charged hadrons.}
\end{figure}

The energy dependence of scaled variance in full phase space is shown in figure~\ref{ed_w}.
An increase of scaled variance with increasing collision energy is observed for p+p, p+n and \Pb collisions. 
For AGS and low SPS energies the scaled variance is smaller than one and the multiplicity distributions are narrower than the corresponding Poisson distributions.
For higher energies the scaled variance is larger than one.
A similar behaviour of p+p and p+n collisions is observed, the small difference is probably caused by the additional proton in p+p collisions,
which does not fluctuate. Therefore the scaled variance for positively and all charged particles is a bit lower in p+p than in p+n collisions.
The scaled variance in \Pb collisions behaves similar as in p+p interactions.
The HSD model yields similar results~\cite{Konchakovski:2007ss}.  

For positively and negatively charged hadrons the scaled variance is similar, where the values are about twice as high for all charged hadrons. This is partly due
to resonances decaying into two oppositely charged particles.
Such a resonance is detected as two charged particles, therefore the fluctuations are increased.

The experimental data on the energy dependence of scaled variance of all charged hadrons in p+p interactions is parametrised in \cite{Heiselberg:2000fk} as
\begin{equation}
\omega(h^\pm) \approx 0.35 \cdot \frac{(<n^\pm>-1)^2}{<n^\pm>}
\end{equation}
The UrQMD results are in agreement with the data at AGS and SPS energies, but the fluctuations are slightly overpredicted at RHIC energies.

Figure~\ref{ed_w} shows that the energy dependence of scaled variance predicted by the UrQMD model is totally different to the predictions of 
a hadron gas model \cite{Begun:2006uu} (for further details the reader is referred also to \cite{Begun:2004gs,Begun:2004pk,Begun:2004zb,Begun:2005ah,Begun:2006jf}).
In the grand-canonical (GCE), canonical (CE) and micro-canonical (MCE) ensemble the scaled variance stays constant for high energies, where in the UrQMD model it strongly
increases with energy.
Therefore experimental data on multiplicity fluctuations, preferably at high (RHIC, LHC) energies, should be able to distinguish between hadron gas and 
string-hadronic models~\cite{Konchakovski:2007ss}.

\begin{figure}
\includegraphics[height=5.9cm]{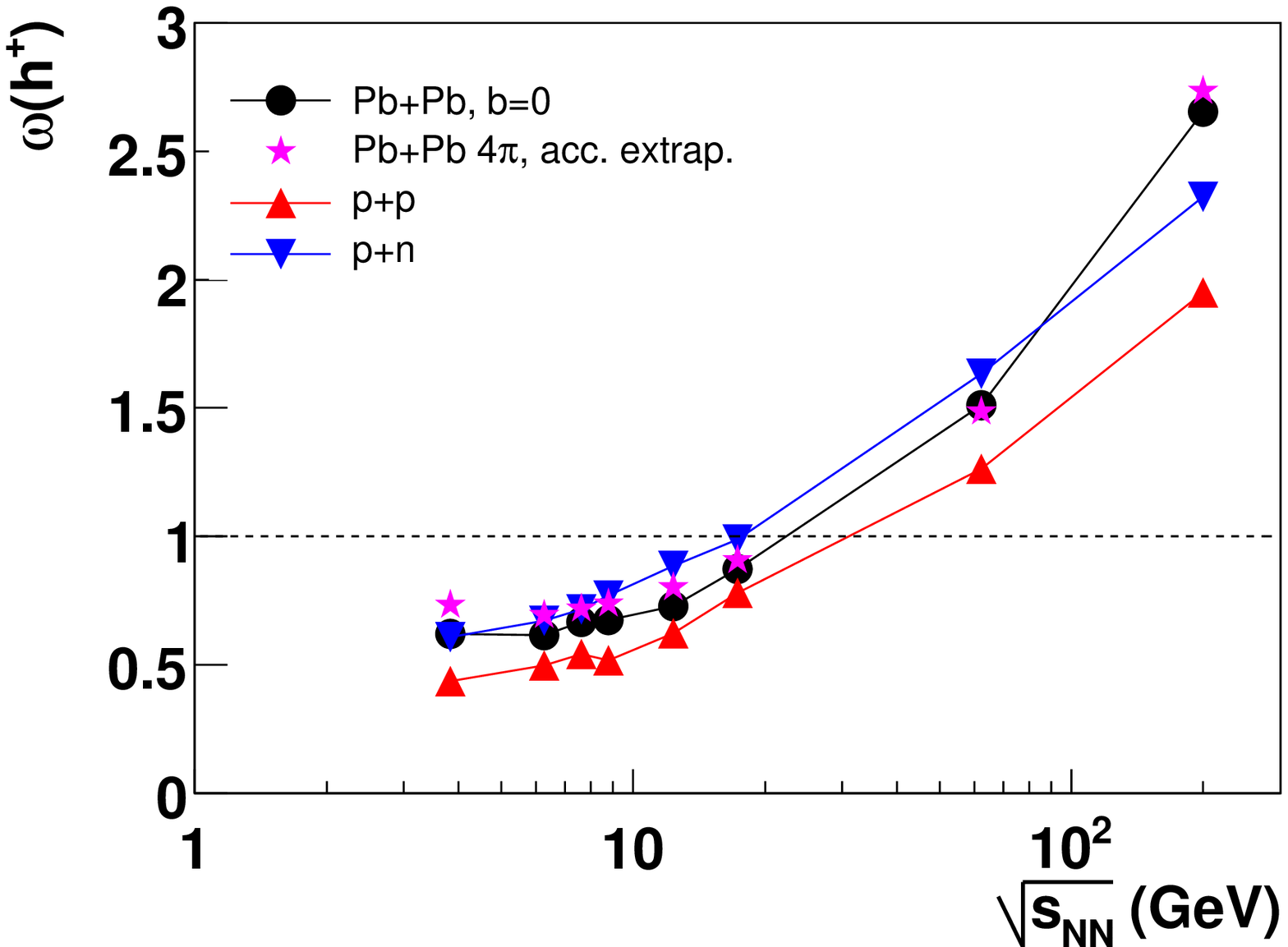}\\
\includegraphics[height=5.9cm]{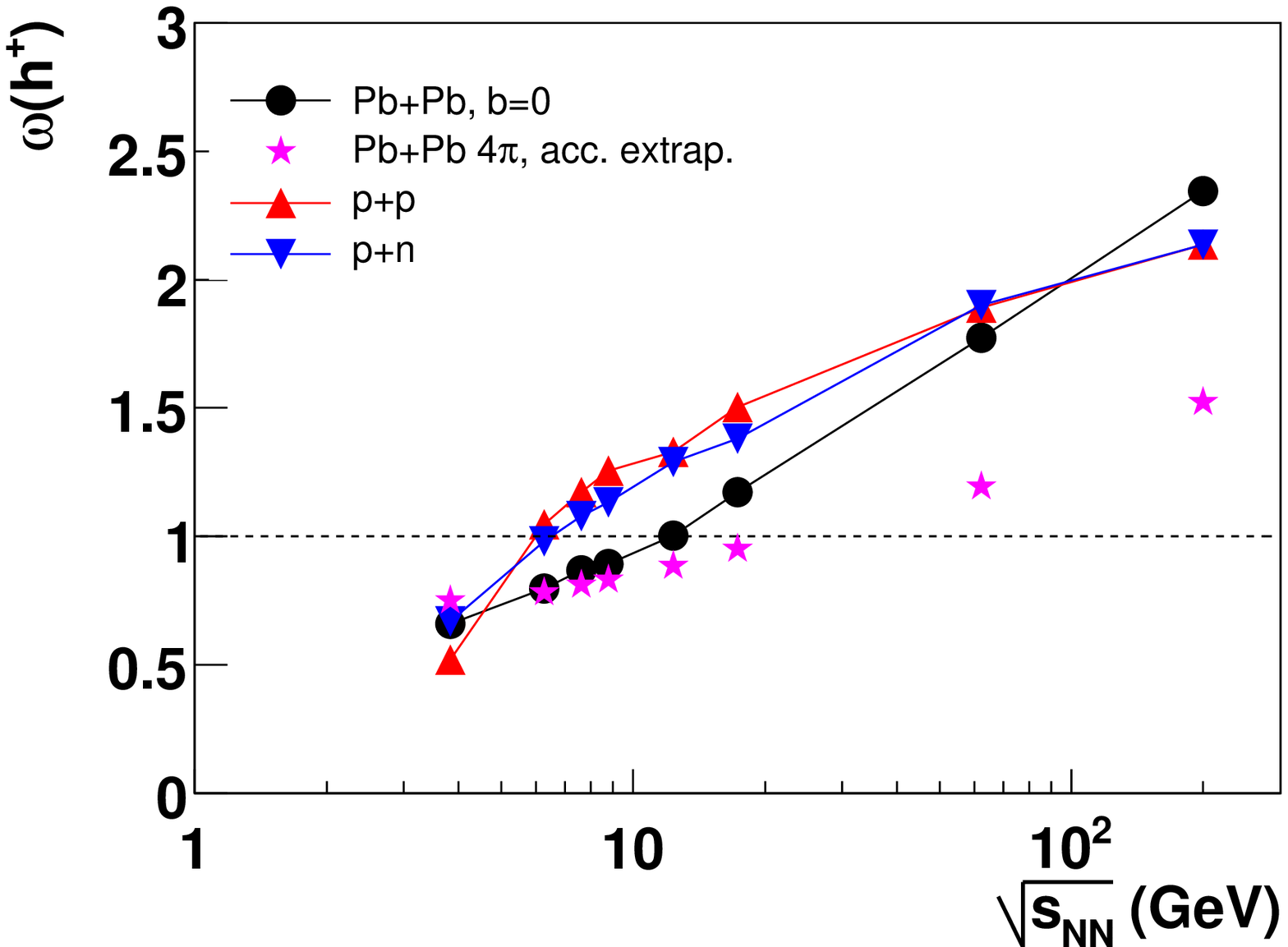}\\
\includegraphics[height=5.9cm]{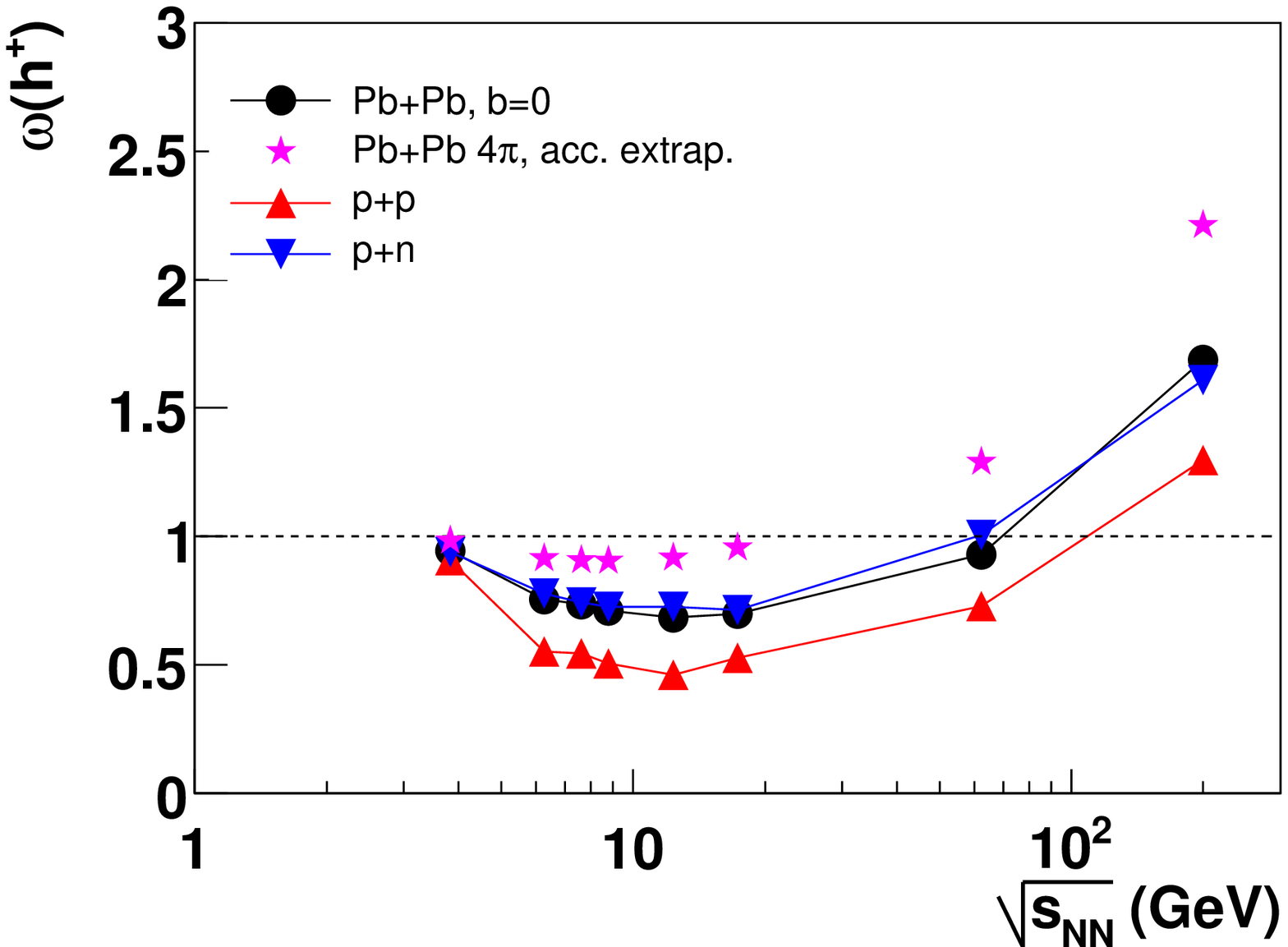}\\
\caption{\label{w_rapi_hp}(Color online) Scaled variance of positively charged hadrons produced in inelastic $p+p$, $p+n$ and central \Pb collisions 
as a function of collision energy. 
Top: $0<y<y_{beam}$, middle: $0<y<1$, bottom: $1<y<y_{beam}$.}
\end{figure}

\begin{figure}
\includegraphics[height=5.9cm]{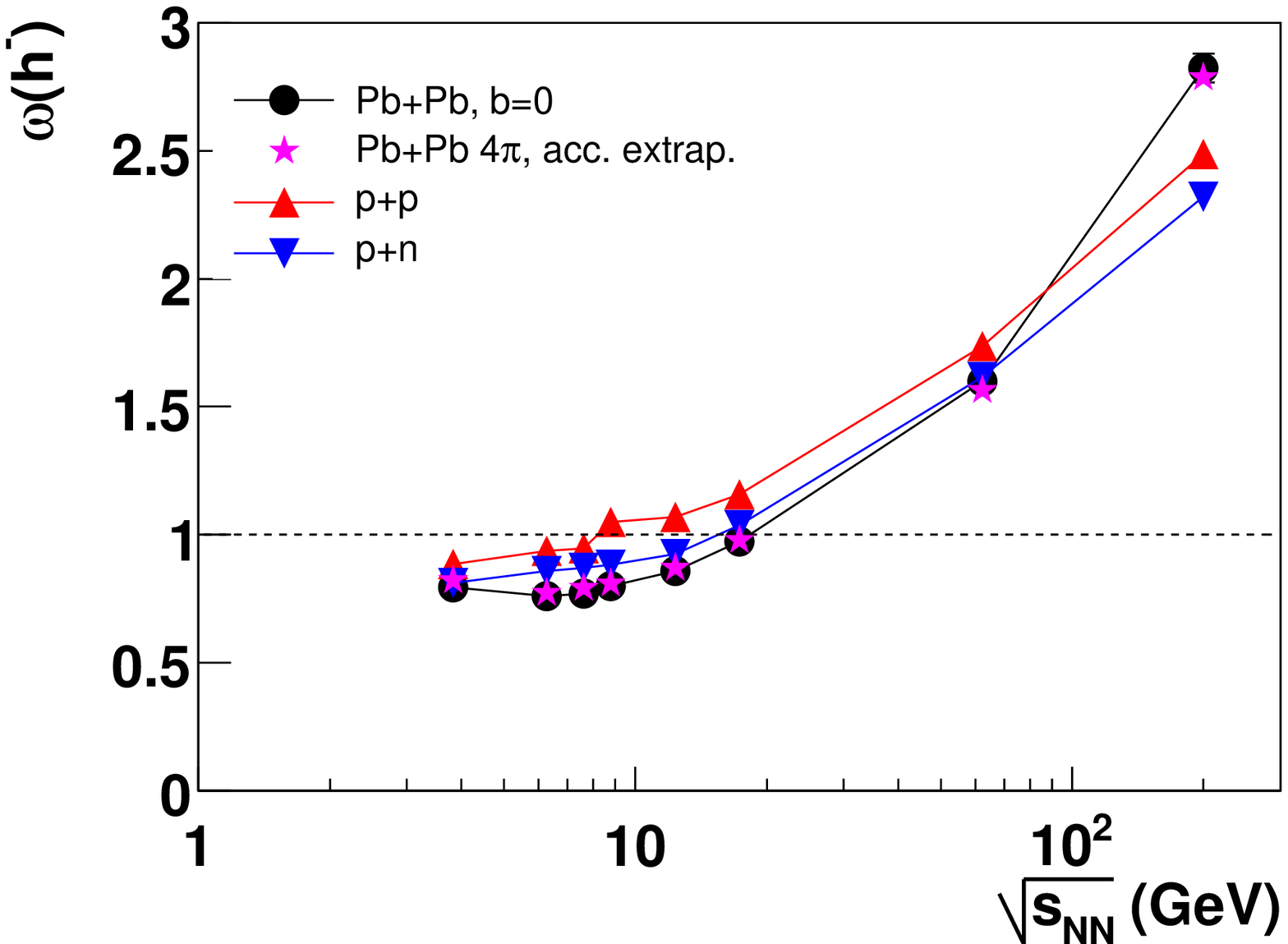}\\
\includegraphics[height=5.9cm]{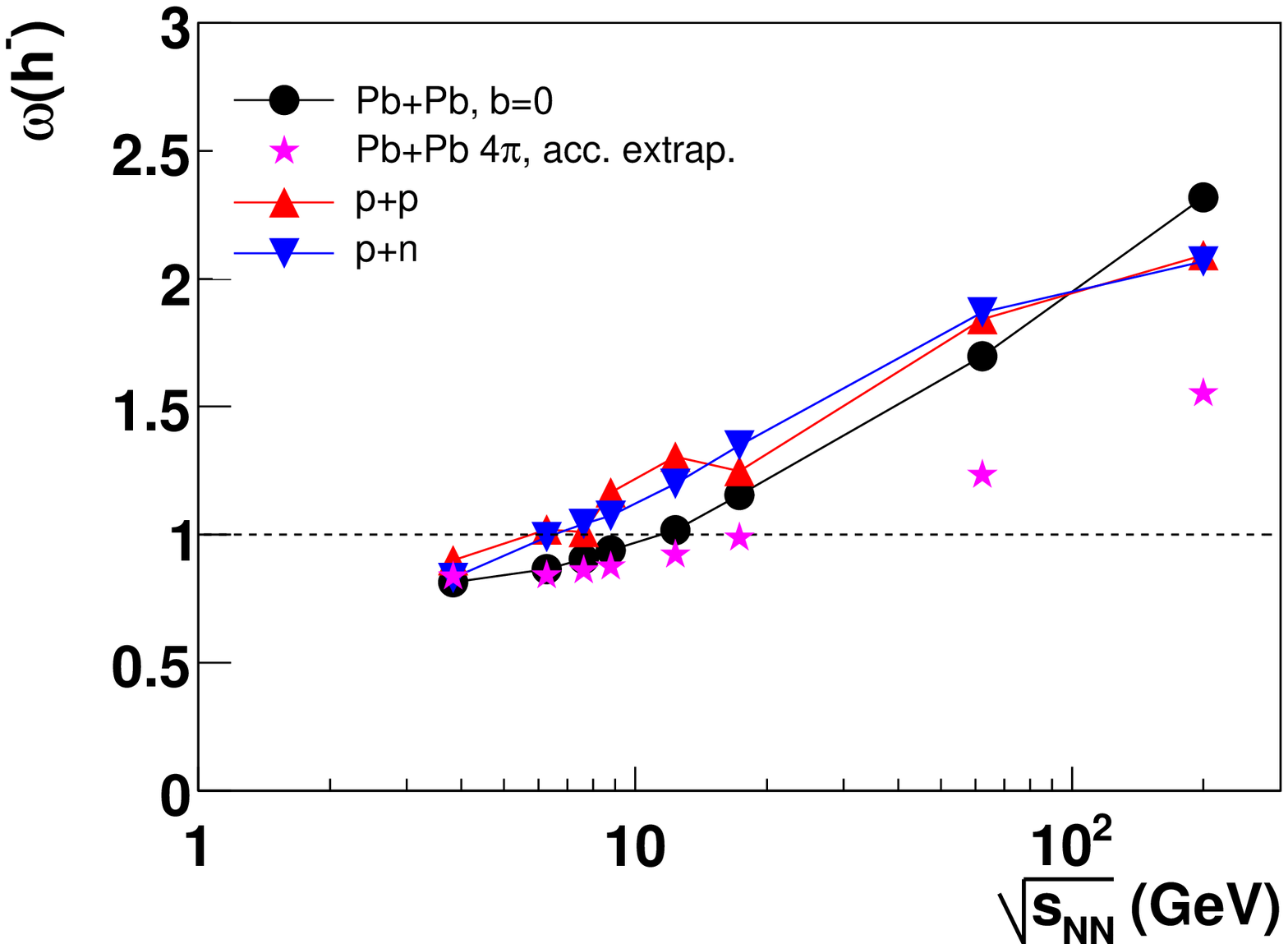}\\
\includegraphics[height=5.9cm]{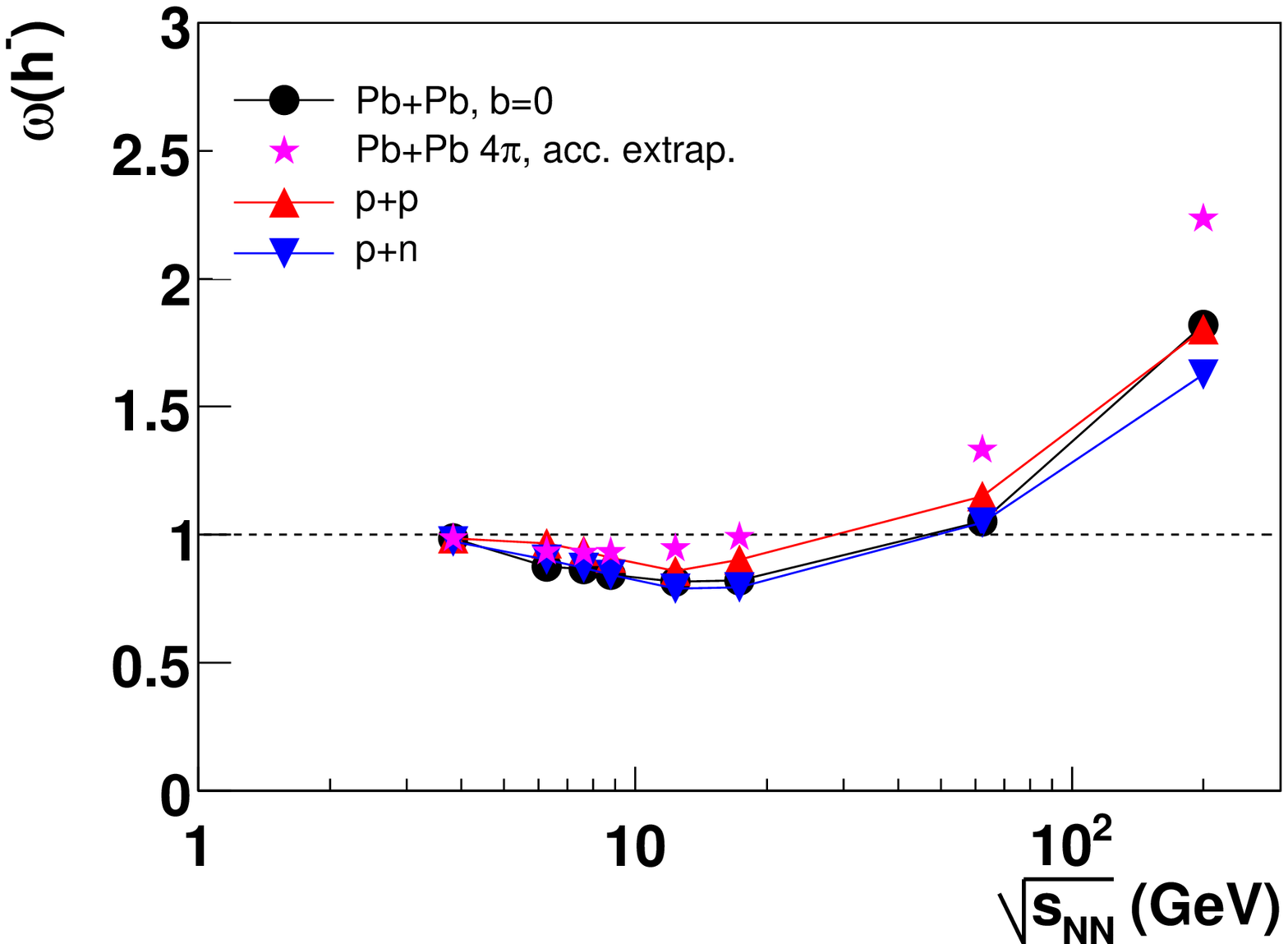}\\
\caption{\label{w_rapi_hm}(Color online) Scaled variance of negatively charged hadrons produced in inelastic $p+p$,
$p+n$ and central \Pb collisions as a function of collision energy. 
Top: $0<y<y_{beam}$, middle: $0<y<1$, bottom: $1<y<y_{beam}$.}
\end{figure}

\begin{figure}
\includegraphics[height=5.9cm]{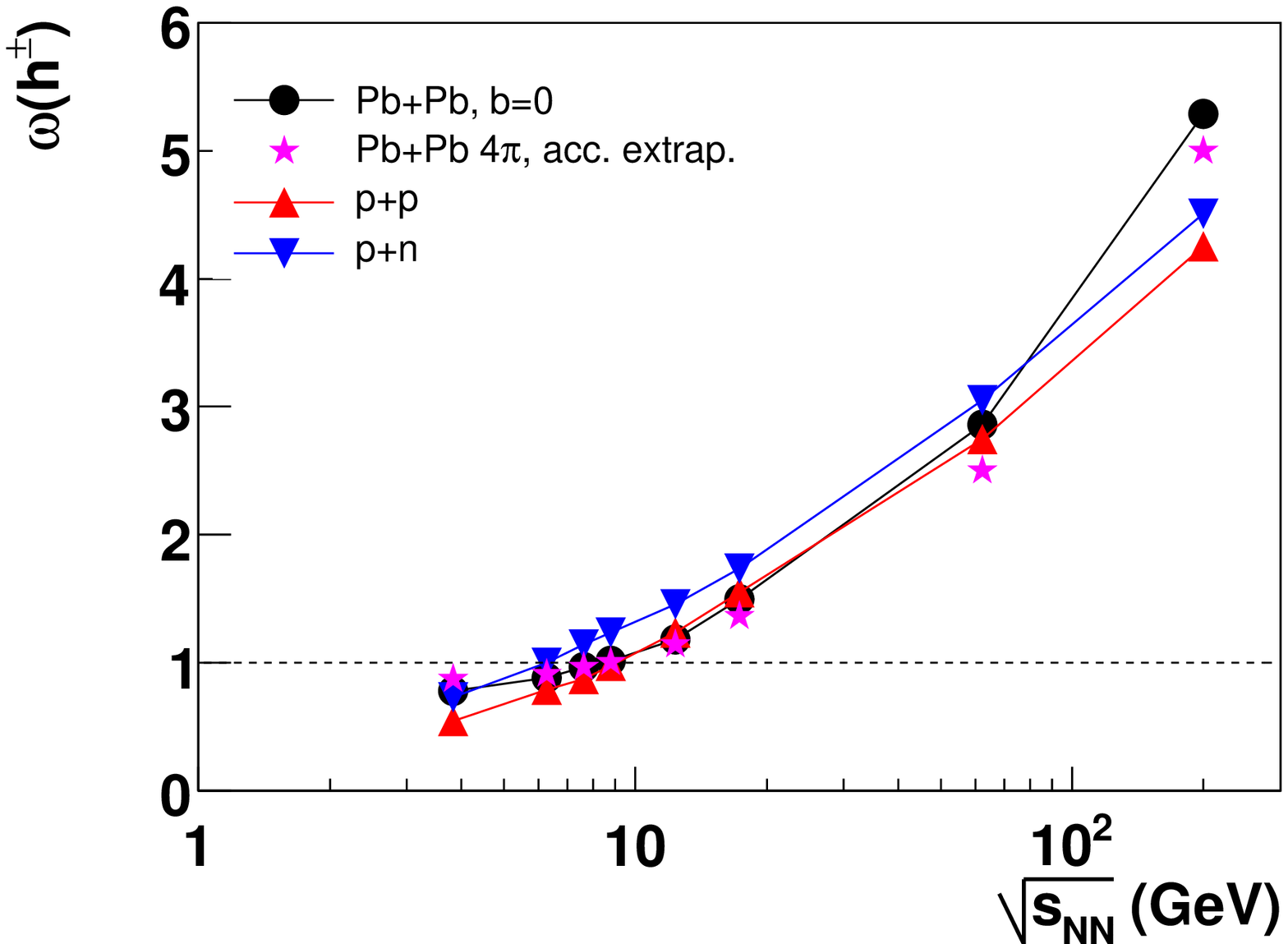}\\
\includegraphics[height=5.9cm]{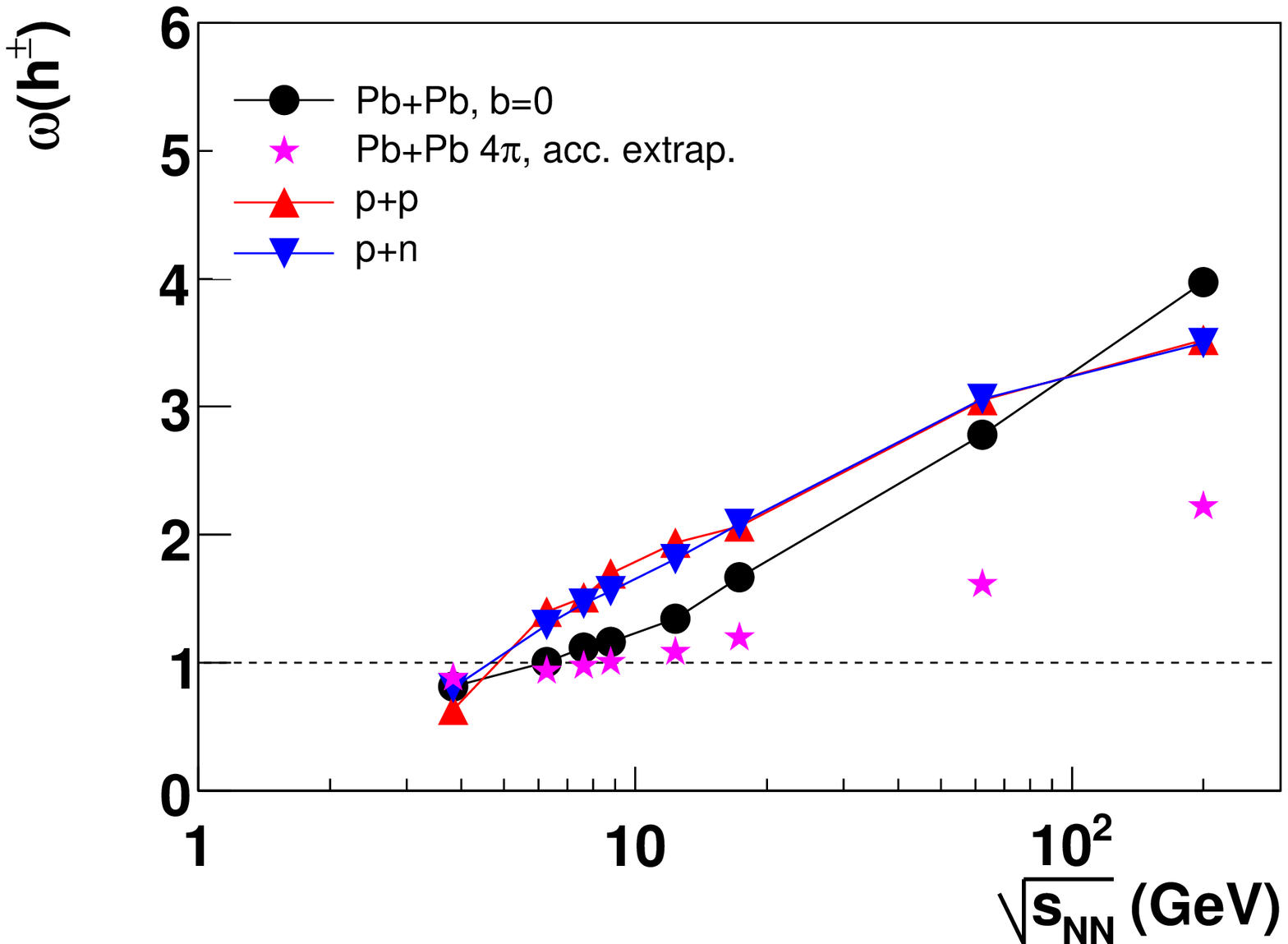}\\
\includegraphics[height=5.9cm]{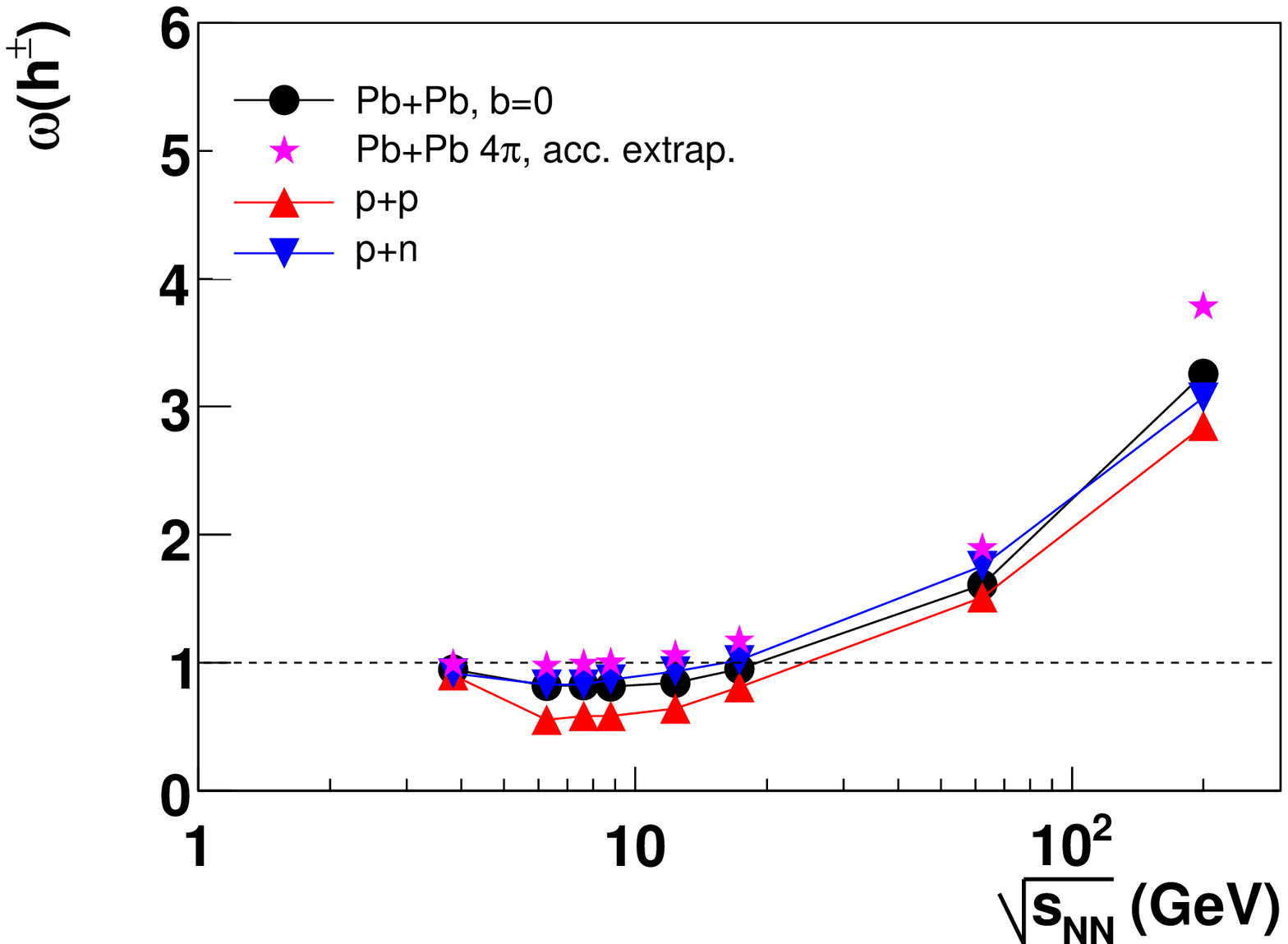}\\
\caption{\label{w_rapi_hpm}(Color online) Scaled variance of all charged hadrons produced in inelastic $p+p$, $p+n$ and central \Pb collisions 
as a function of collision energy. 
Top: $0<y<y_{beam}$, middle: $0<y<1$, bottom: $1<y<y_{beam}$.}
\end{figure}

For a more differential study of fluctuations and for a better comparison to experimental results, 
three different rapidity intervals, one at midrapidity $0<y<1$,
one at forward rapidity $1<y<y_{beam}$ and a combination of both $0<y<y_{beam}$, covering most of the forward hemisphere,
were taken. For $p+n$ collisions the forward hemisphere includes the rapidity of the projectile neutron.
The scaled variance for these intervals for
positively, negatively and all charged particles are shown in figures \ref{w_rapi_hp}-\ref{w_rapi_hpm}.

As in full phase space, in the three different rapidity intervals a similar behaviour of scaled variance of $p+p$, $p+n$ and \Pb collisions was observed.

The energy dependence of fluctuations in the forward hemisphere ($0<y<y_{beam}$) looks similar to the one in the full phase space, the absolute number of 
scaled variance is similar to the result expected when applying the acceptance extrapolation according 
to formula~\ref{wscale} (shown as stars in figures~\ref{w_rapi_hp}-\ref{w_rapi_hpm}).

For low energies a large fraction of particles is in the midrapidity interval ($0<y<1$) where a very small amount of particles is in the forward rapidity interval
($1<y<y_{beam}$). With increasing energy both the width and the number of particles in the forward 
rapidity interval increases strongly, where the number of particles in the forward rapidity interval 
increases only weakly.

At midrapidity ($0<y<1$) the scaled variance is in the same order of magnitude as in the rapidity interval 
$0<y<y_{beam}$, but the mean multiplicity is much lower.
The acceptance extrapolation formula~\ref{wscale} strongly underpredicts fluctuations in this rapidity region.
At forward rapidity ($1<y<y_{beam}$) the fluctuations are much smaller than predicted by the acceptance extrapolation
formula. 
For lower energies the scaled variance decreases with energy, for higher energies it increases. This can be qualitatively understood by the interplay of an increasing
fraction of particles in this rapidity interval and an increasing scaled variance in $4\pi$, which is smaller than $1$ for lower and larger than $1$ for 
higher energies.

\section{Rapidity and Transverse Momentum Dependence}

\begin{figure}[h]
\includegraphics[height=5.9cm]{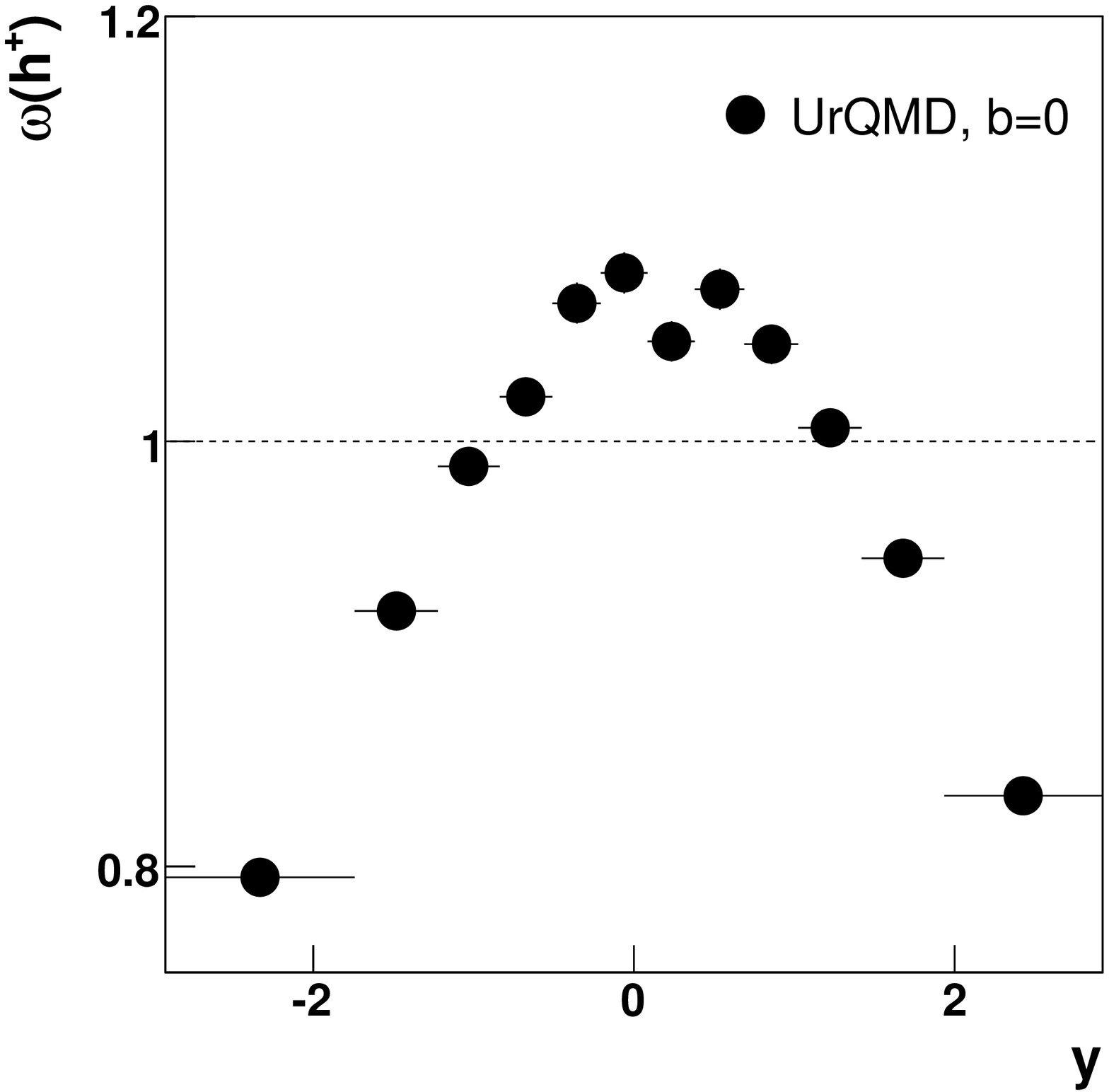}
\includegraphics[height=5.9cm]{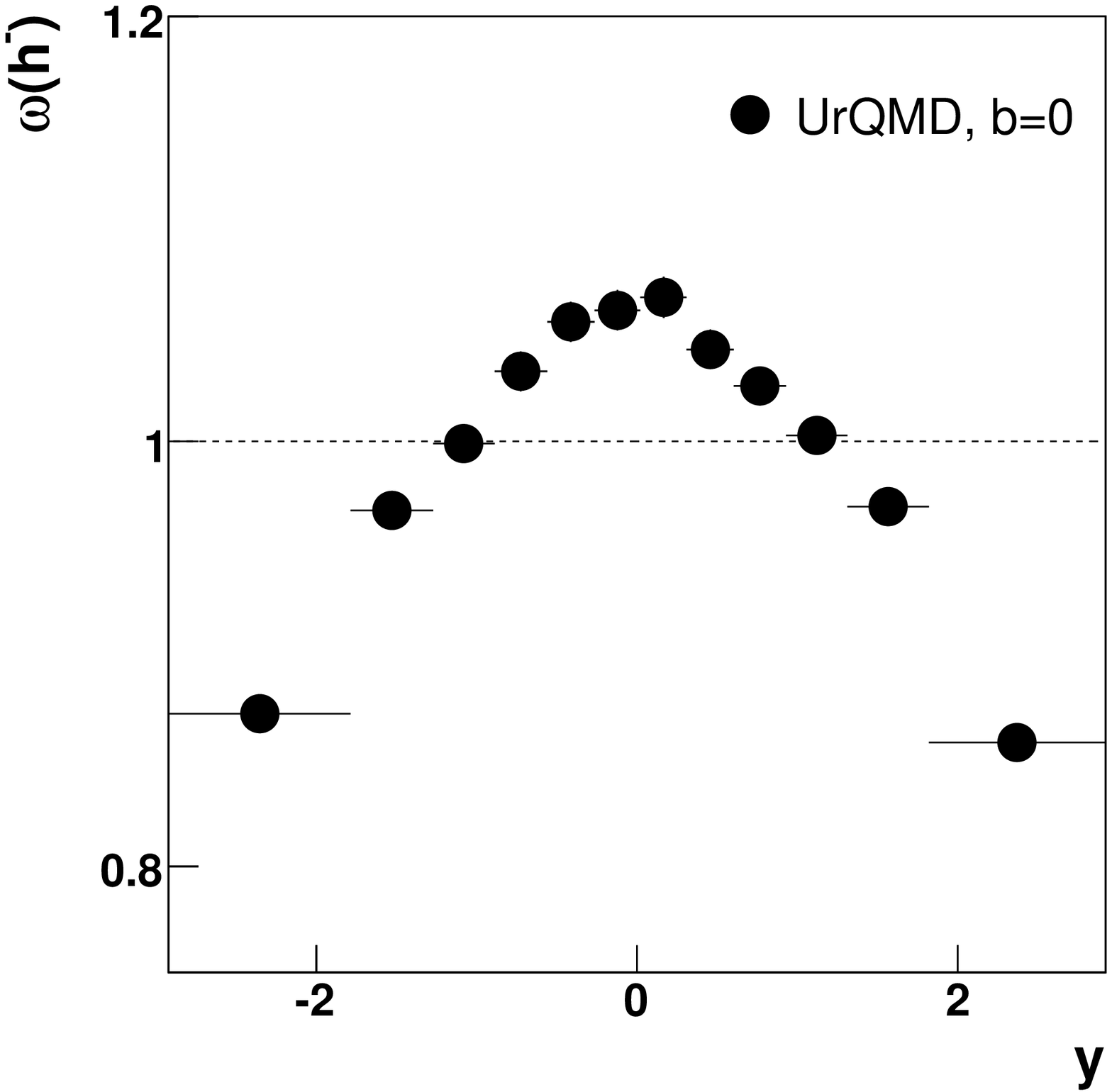}
\includegraphics[height=5.9cm]{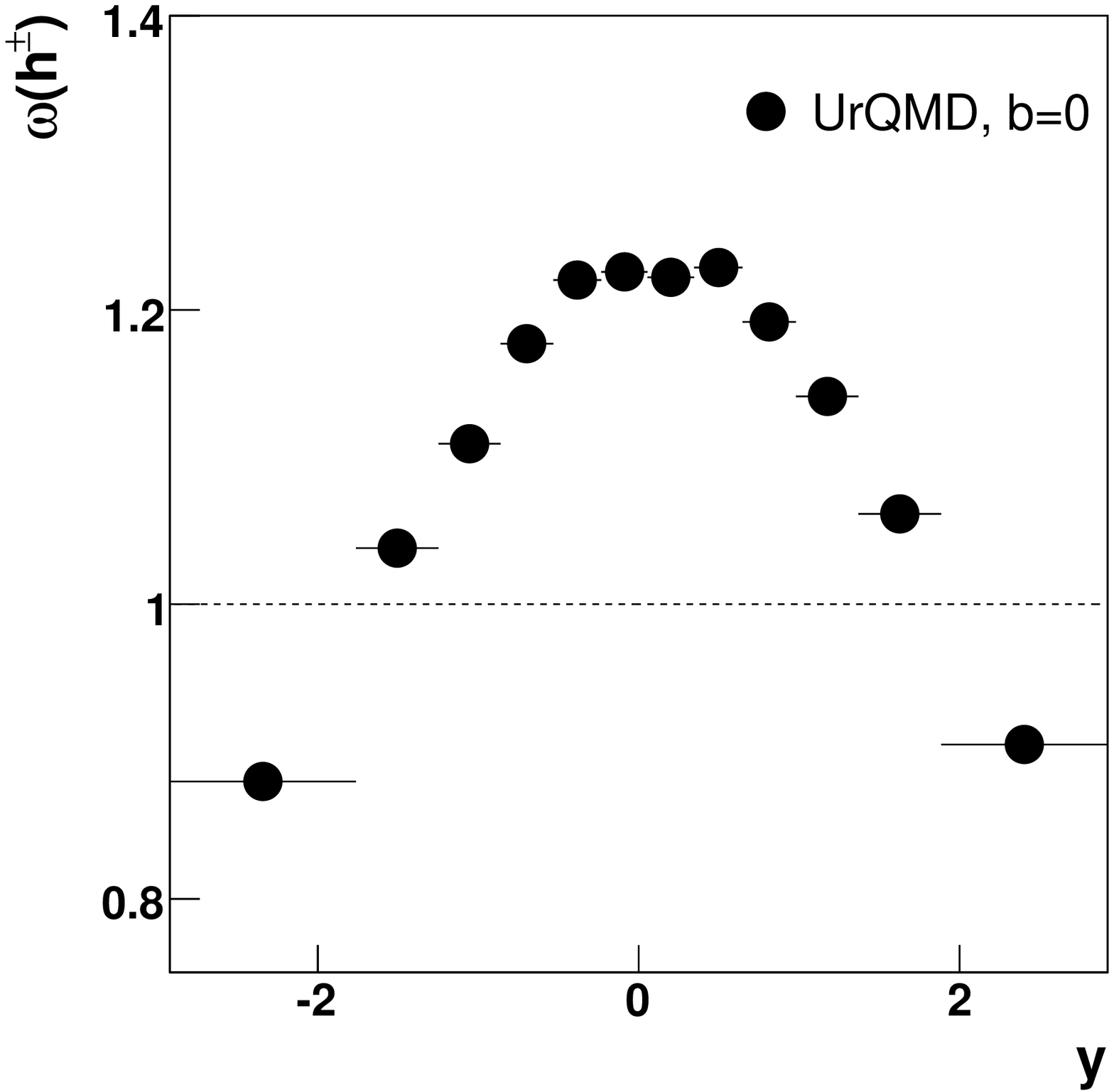}
\caption{\label{ydep_4pi}Rapidity dependence of scaled variance in UrQMD simulation performed in full acceptance of positive (top), 
negative (middle) and all charged (bottom)
hadrons in central \Pb collisions at \eh. The rapidity bins are constructed in such 
a way that the mean multiplicity in each bin is the same.
}
\end{figure}

As already showed in section ~\ref{ed_s}, the scaled variance is a non-trivial function of the selected phase-space. In order to study the dependence 
of scaled variance on rapidity, 12 different rapidity intervals are constructed in such a way that the mean multiplicity in each interval is the same.
If the scaled variance would follow the acceptance scaling formula~\ref{wscale}, the scaled variance would be the same in each interval. In figure~\ref{ydep_4pi}
it is shown that this is not the case. The scaled variance is much higher near midrapidity than in forward and backward rapidities. 

\begin{figure}[h]
\includegraphics[height=5.9cm]{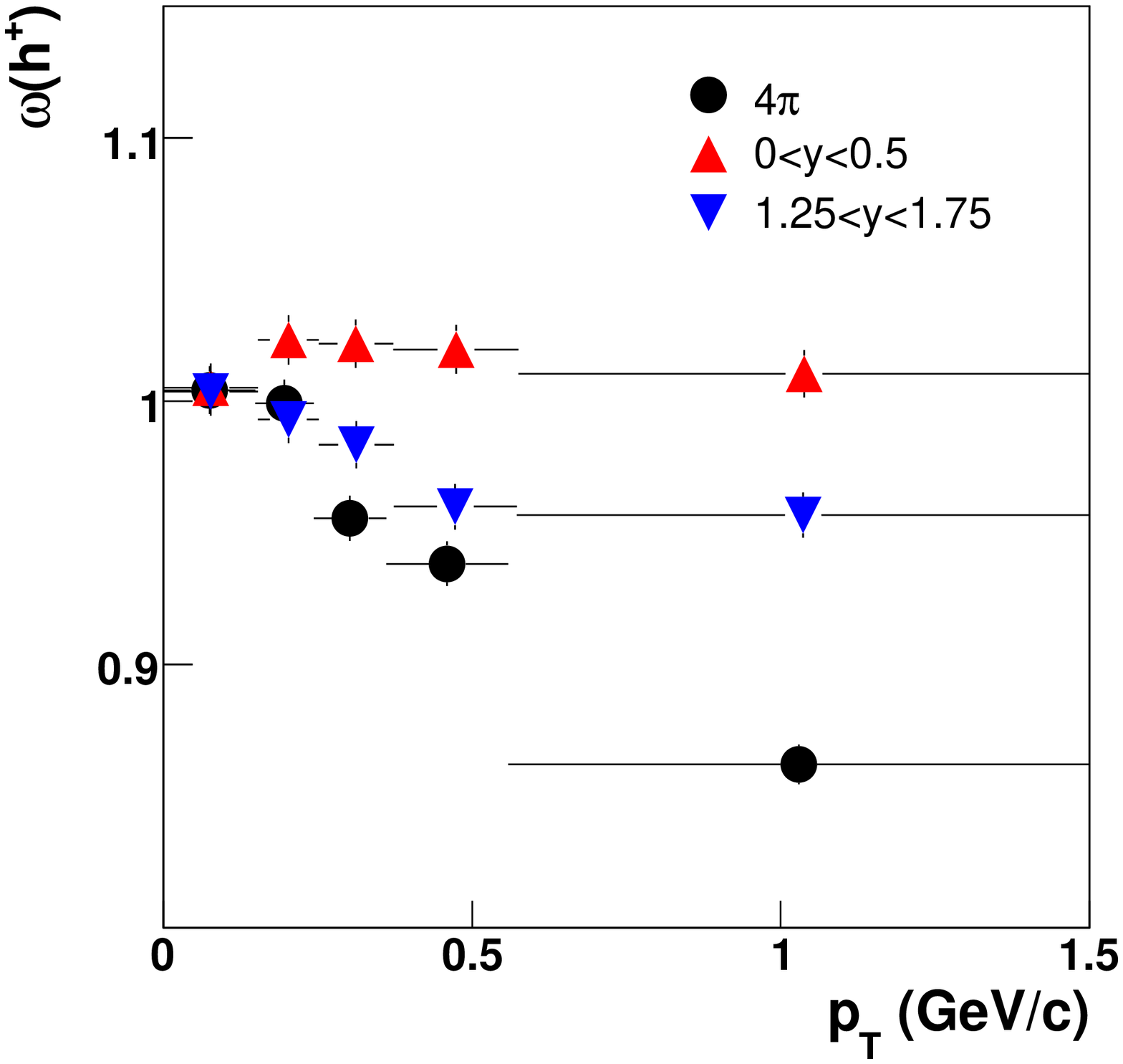}
\includegraphics[height=5.9cm]{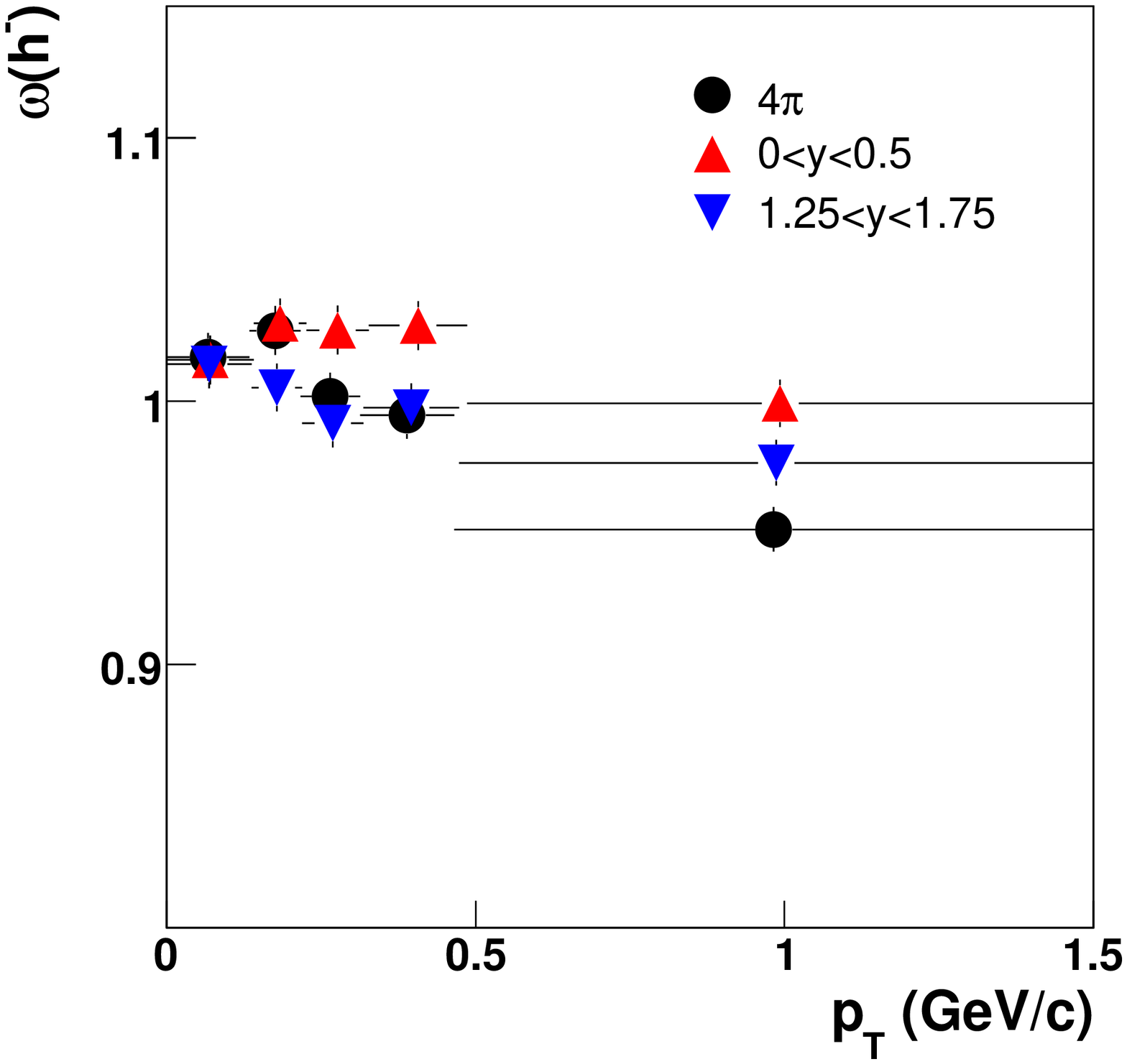}
\includegraphics[height=5.9cm]{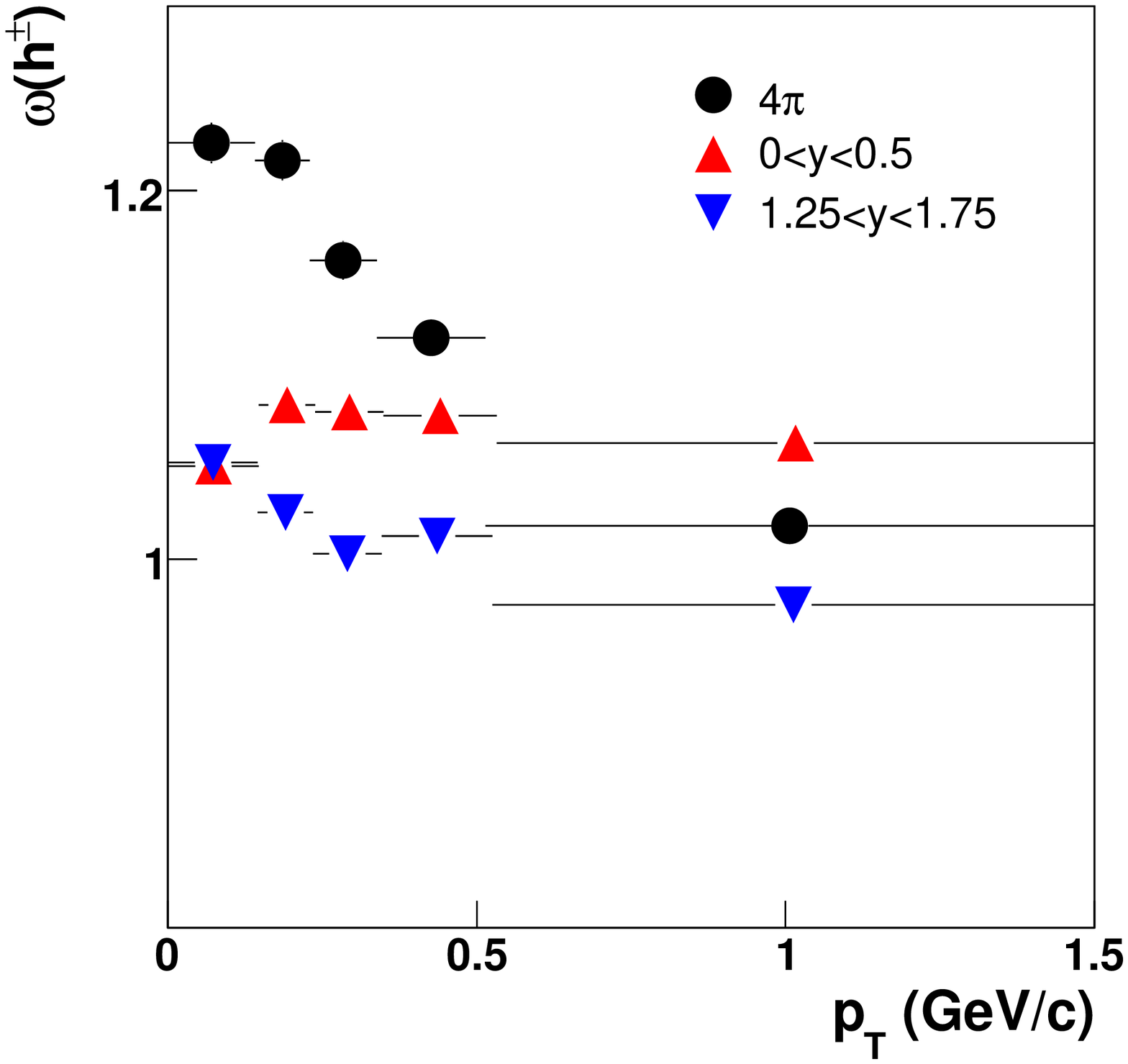}
\caption{\label{ptdep}(Color online) Transverse momentum dependence of multiplicity fluctuations of positively (top), negatively (middle) and all charged
hadrons (bottom) for all rapidities), $0<y<0.5$ and $1.25<y<1.75$.
The transverse momentum bins are constructed in such 
a way that the mean multiplicity in each bin is the same.}
\end{figure}

The transverse momentum dependence of scaled variance is shown in figure~\ref{ptdep} for the full longitudinal phase-space and for a midrapidity and a 
forward rapidity interval. 
The scaled variance decreases with increasing transverse momentum for the full acceptance and at forward rapidity. At midrapidity it stays approximately constant.
The decrease of scaled variance is stronger for positively charged hadrons than for negatively charged ones because the protons, which have smaller relative fluctuations
due to the large number of protons which enter the collision, have a larger mean transverse momentum.

A similar effect of decreasing fluctuations for larger rapidities and transverse momenta
is observed as a result of energy- and momentum conservation in a hadron gas model using the micro-canonical ensemble~\cite{Hauer:ip}.
It costs more energy to create a particle with high momentum,
therefore their number is expected to fluctuate less.

\begin{figure}[h]
\includegraphics[height=5.9cm]{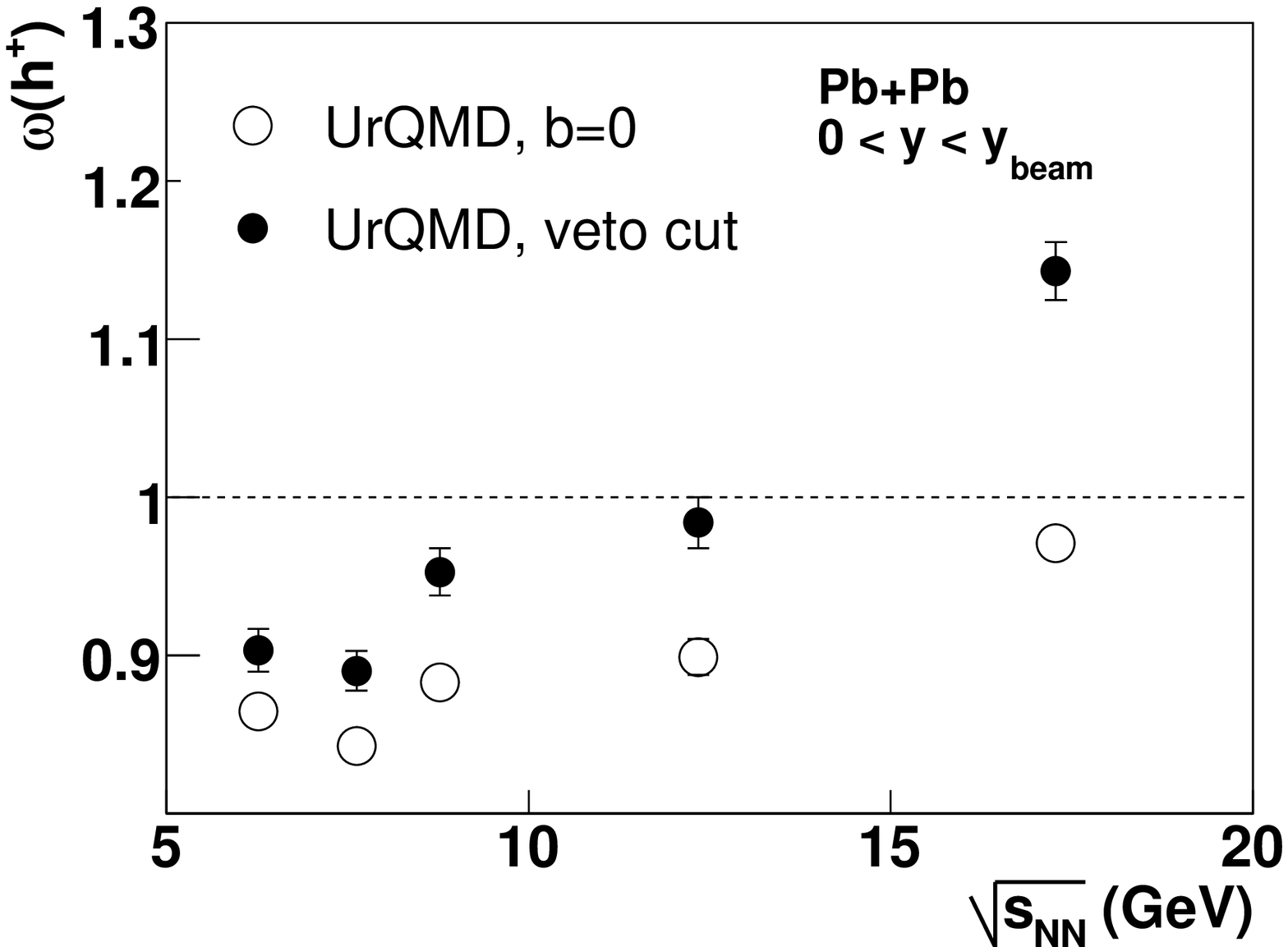}
\includegraphics[height=5.9cm]{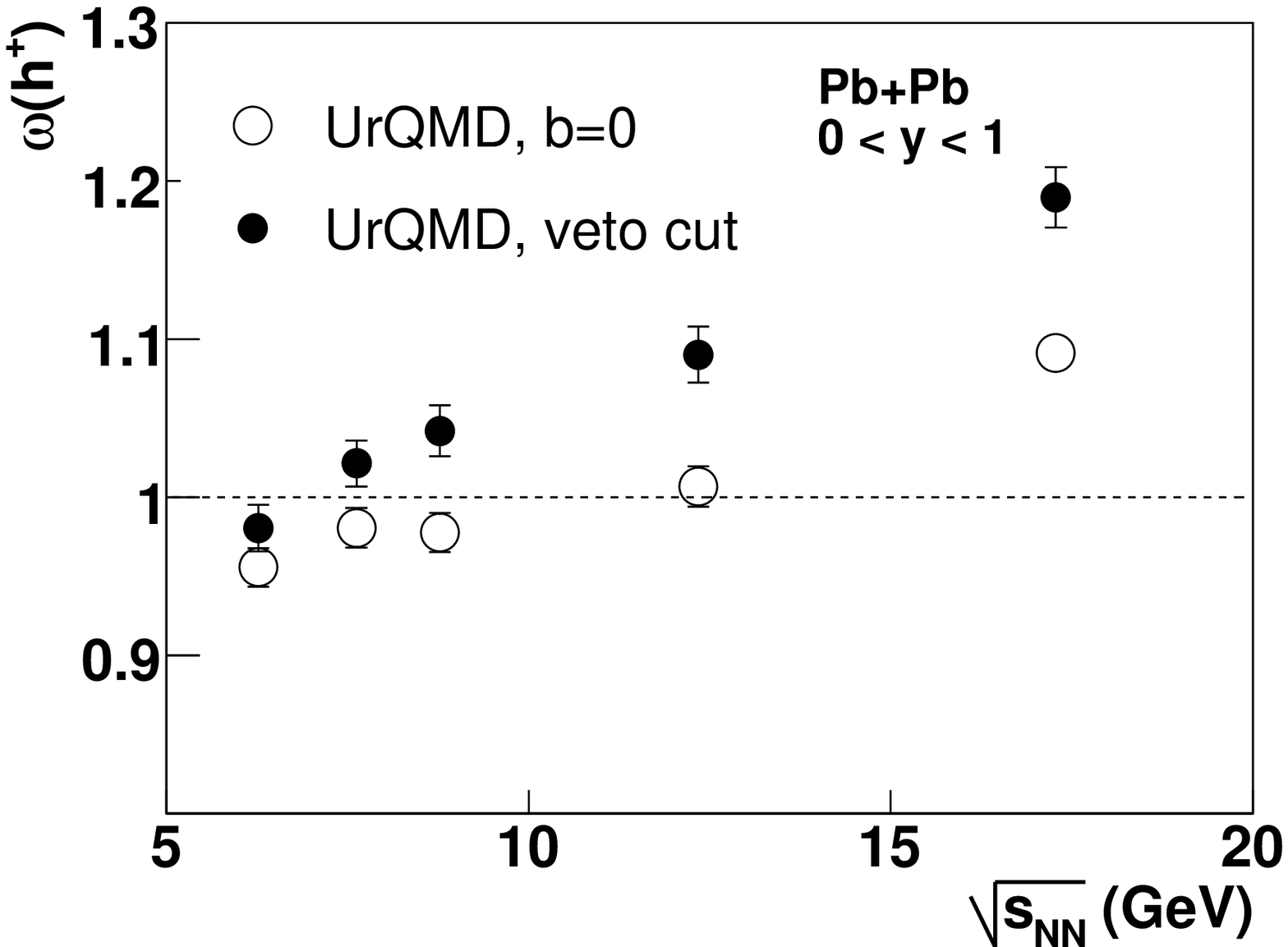}\\
\includegraphics[height=5.9cm]{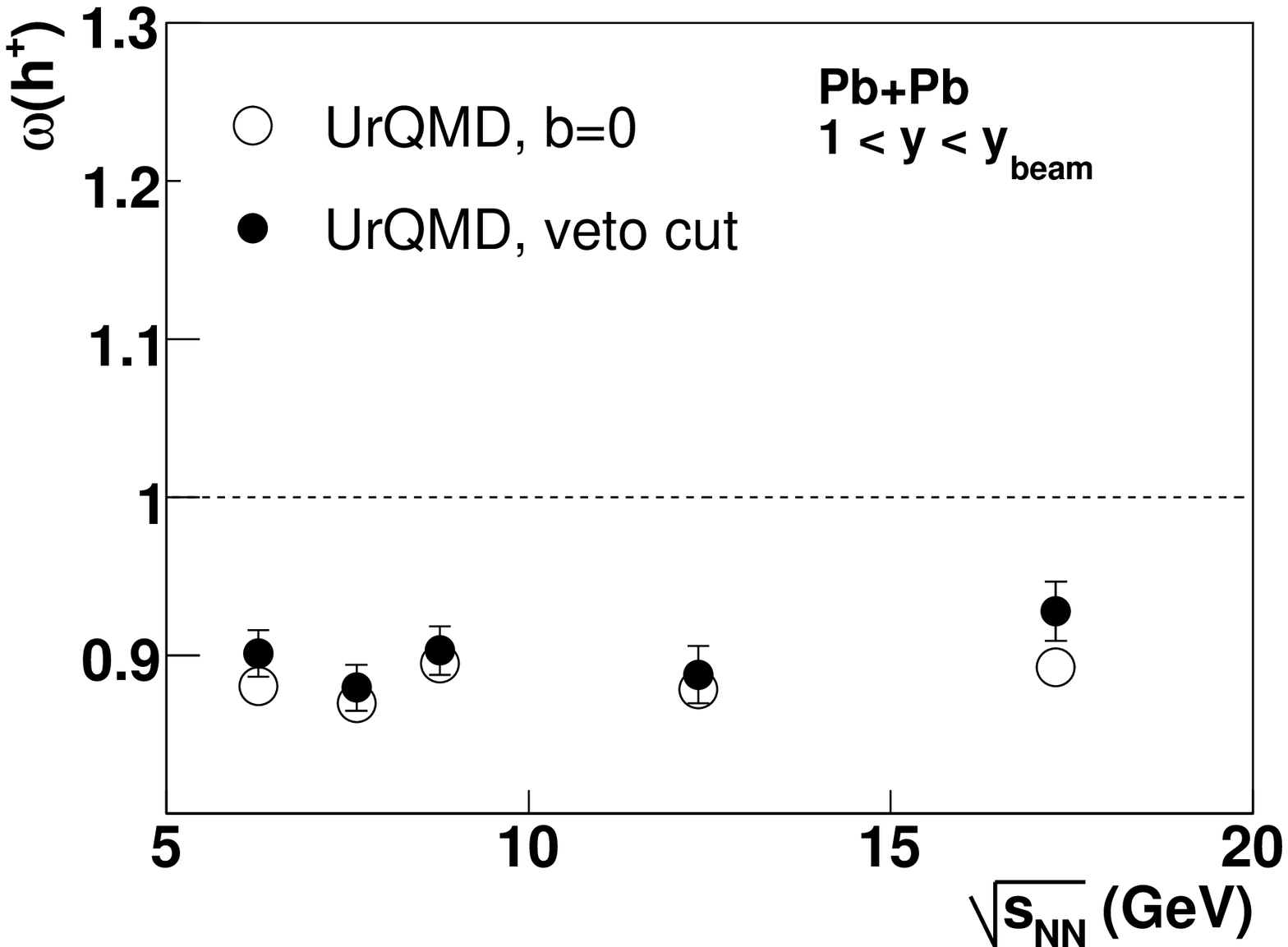}\\
\caption{\label{w_hp}UrQMD predictions for scaled variance of positively charged hadrons in NA49 acceptance produced in central \Pb collisions as a 
function of collision energy. 
Top: $0<y<y_{beam}$, middle: $0<y<1$, bottom: $1<y<y_{beam}$.}
\end{figure}

\begin{figure}[h]
\includegraphics[height=5.9cm]{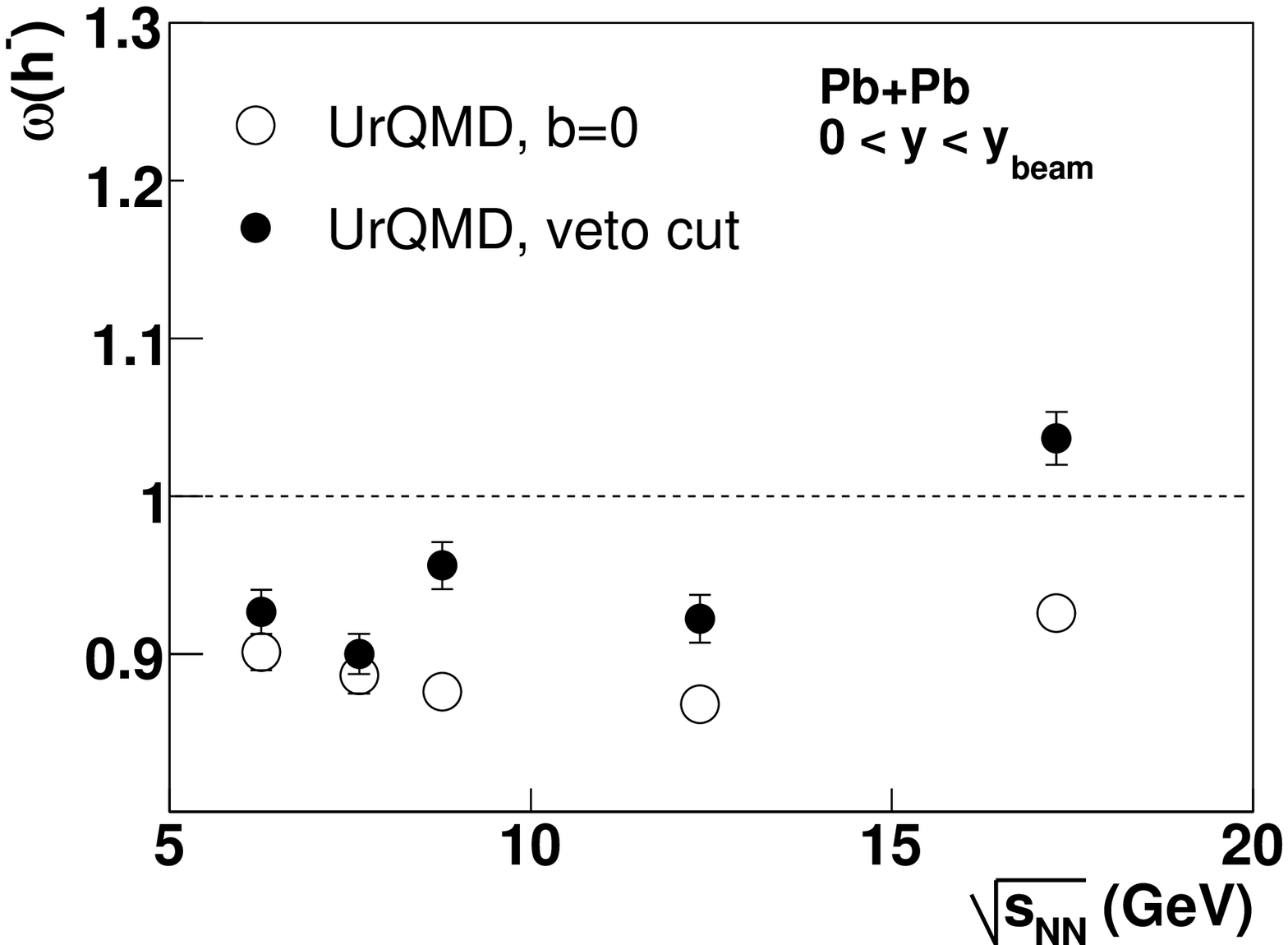}
\includegraphics[height=5.9cm]{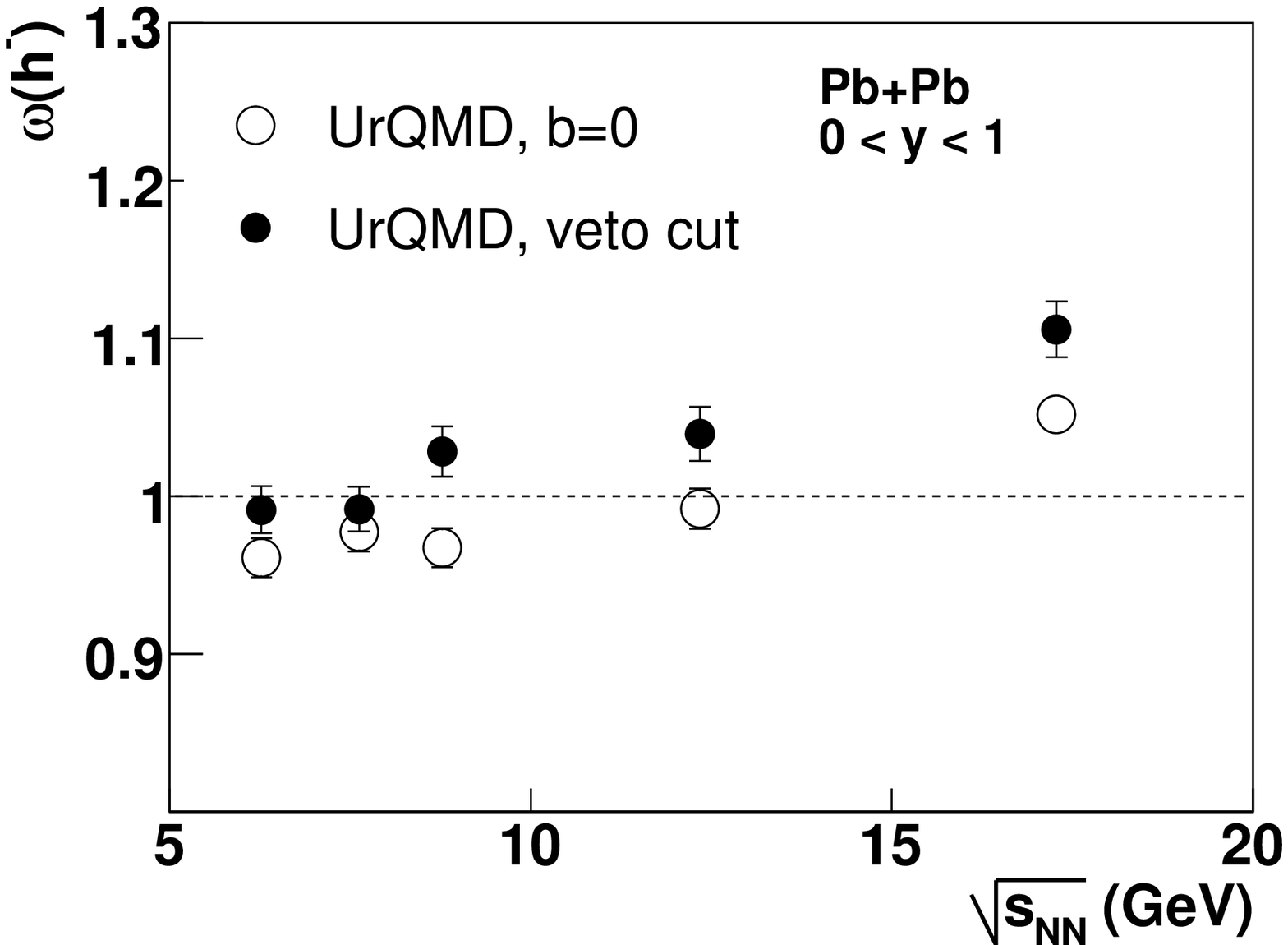}\\
\includegraphics[height=5.9cm]{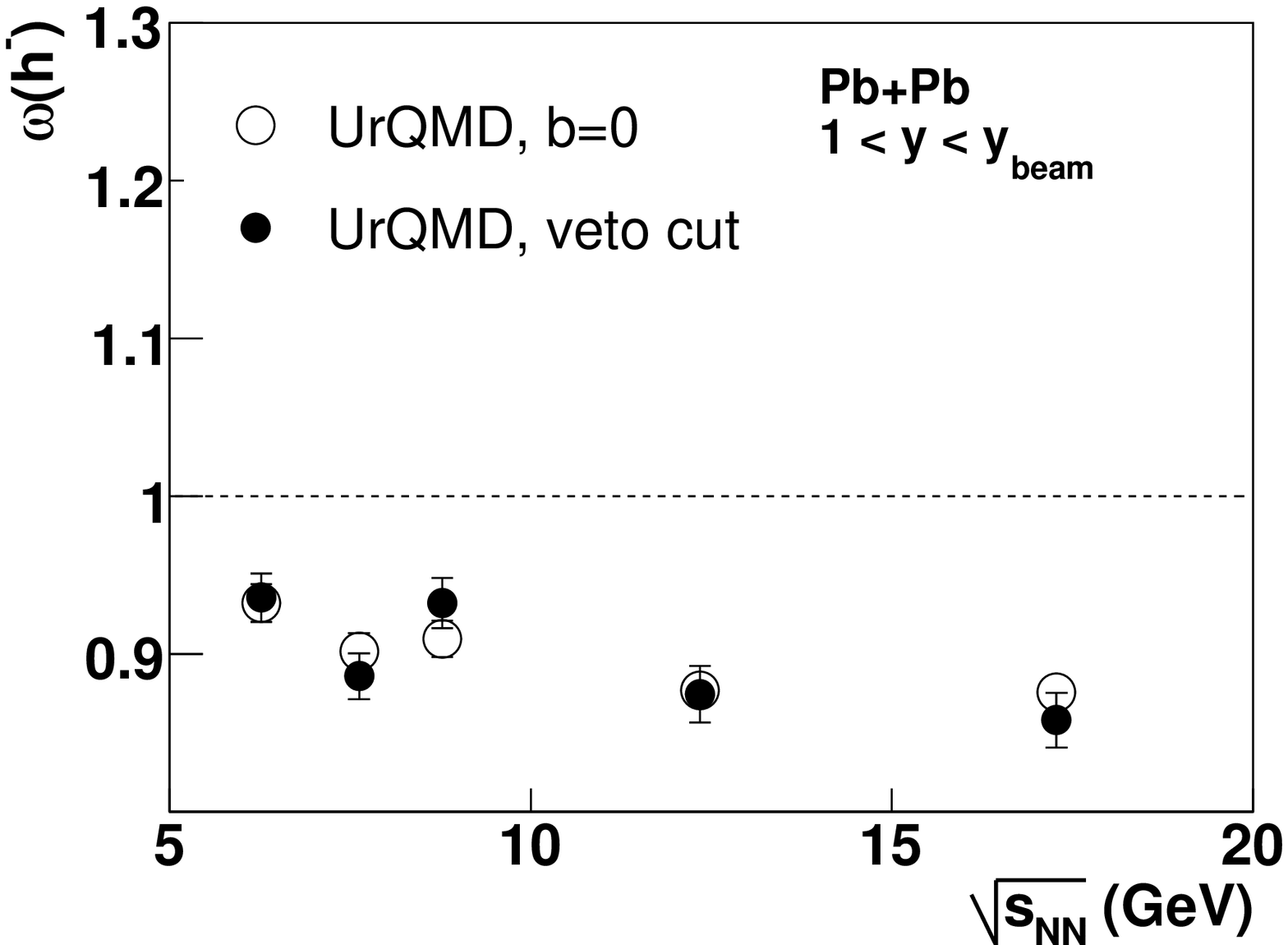}\\
\caption{\label{w_hm}UrQMD predictions for scaled variance of negatively charged hadrons in NA49 acceptance produced in central \Pb collisions as a 
function of collision energy. 
Top: $0<y<y_{beam}$, middle: $0<y<1$, bottom: $1<y<y_{beam}$.}
\end{figure}

\begin{figure}[h]
\includegraphics[height=5.9cm]{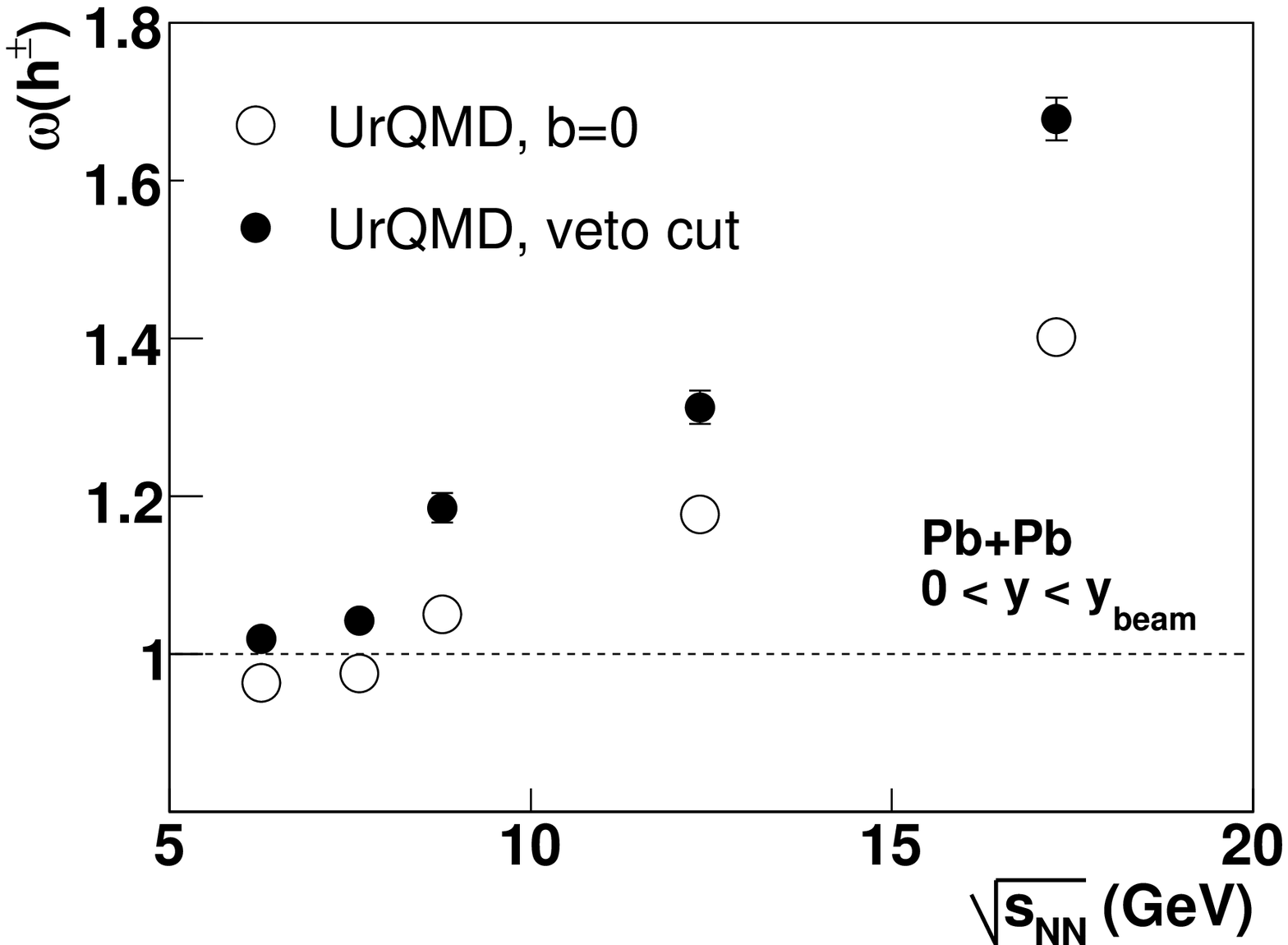}
\includegraphics[height=5.9cm]{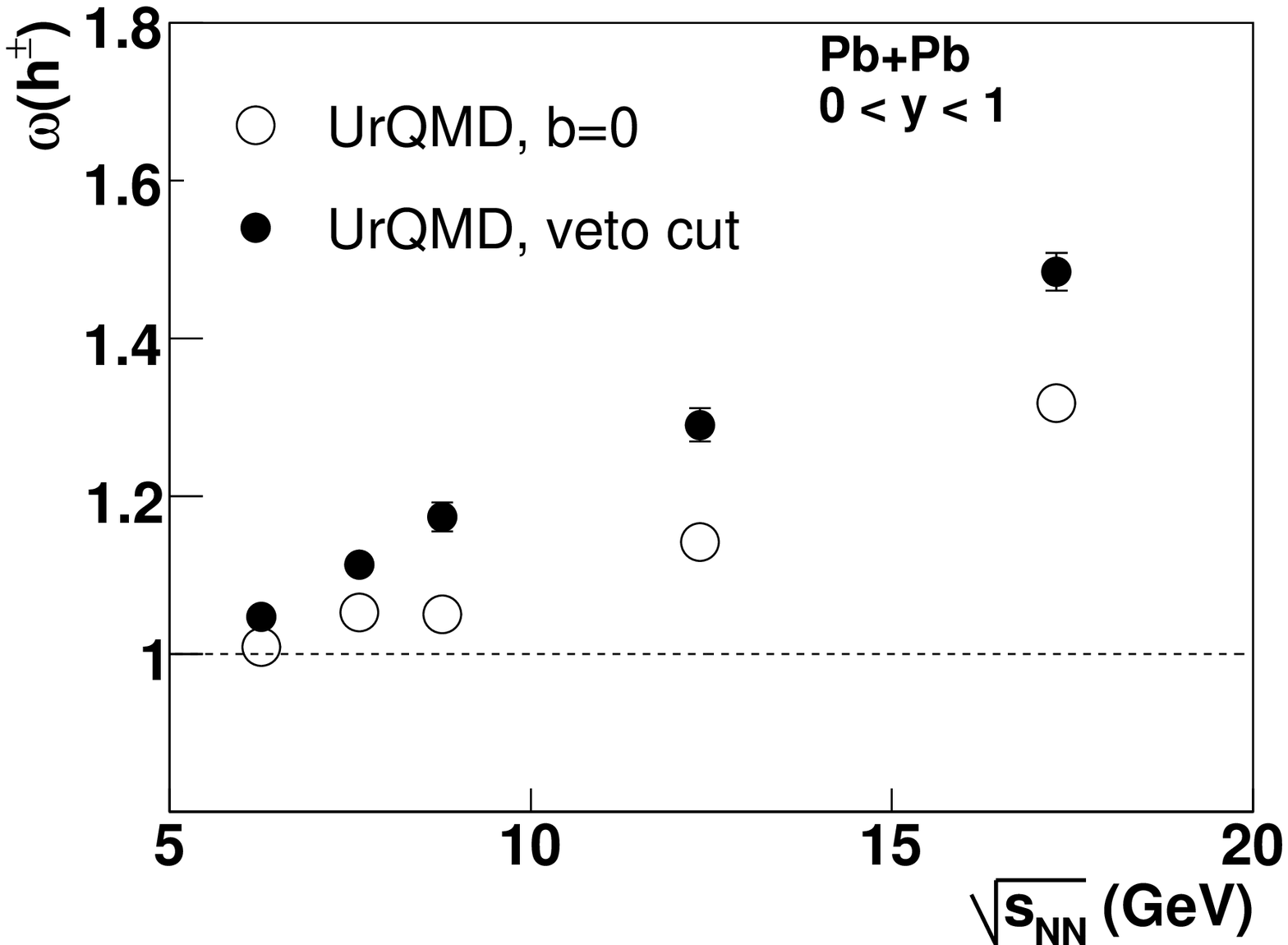}\\
\includegraphics[height=5.9cm]{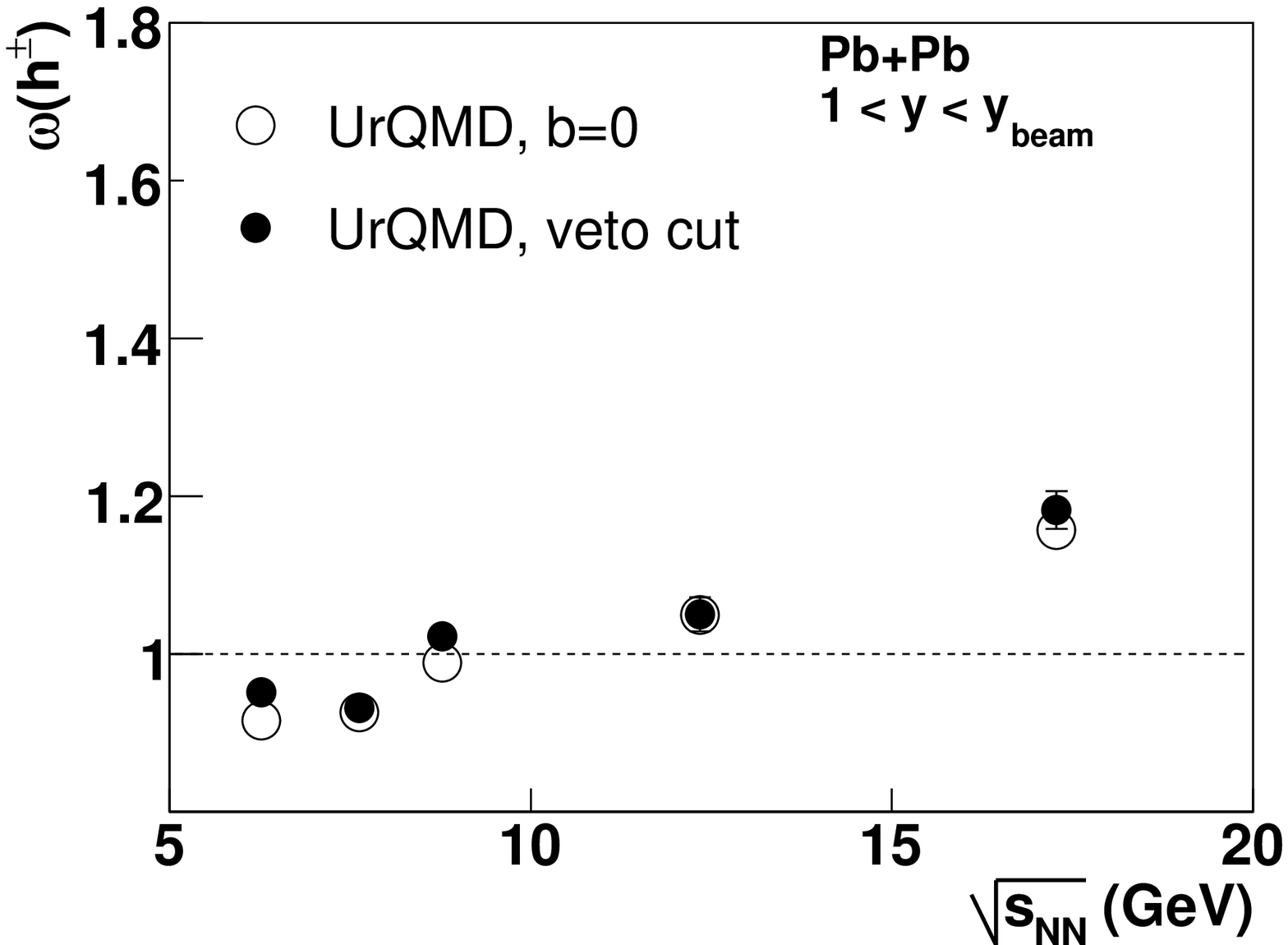}\\
\caption{\label{w_hpm}UrQMD predictions for scaled variance of all charged hadrons in NA49 acceptance produced in central \Pb collisions as a 
function of collision energy. 
Top: $0<y<y_{beam}$, middle: $0<y<1$, bottom: $1<y<y_{beam}$.}
\end{figure}

\section{Predictions for the NA49 Experiment}

Preliminary data of the NA49 experiment on the energy dependence of multiplicity fluctuations in very central \Pb collisions was shown 
in~\cite{Lungwitz:2006cx,Lungwitz:2006cy}.
Final data obtained in a larger geometrical acceptance will be published soon. 

In order to compare the experimental data with model calculations, both the geometrical acceptance of the detector and the centrality selection have to be
implemented in the model calculation.

The geometrical acceptance of the NA49 experiment~\cite{Afanasev:1999iu} is located mostly in the forward hemisphere. The acceptance defined by the detector
geometry and the track selection criteria is different for each collision energy and a complicated function of the particle momentum $\vec{p}$.
Acceptance tables in $y(\pi)$, $p_T$ and $\phi$ can be obtained at the author`s website (http://www.ikf.physik.uni-frankfurt.de/users/lungwitz/acceptance/). 
In the NA49 experiment it is not possible to identify a particle on the track-by-track basis, therefore for the calculation of rapidity in the
fixed target laboratory system pion mass is assumed. The rapidity is then transformed into the center of mass system of the collision.\\
For the UrQMD model predictions showed in this section both the geometrical acceptance defined by the acceptance tables and the assumption of pion mass when
calculating rapidity and transforming into the center of mass system are taken into account.

In the NA49 experiment the centrality of a collision can be measured by the energy of projectile spectators, which are registered by a calorimeter.
This veto calorimeter is adjusted in such a way that it registers all spectator protons, neutrons and fragments of the projectile. The lower the energy in the veto 
calorimeter the more central is the collision. For the multiplicity fluctuation analysis the $1\%$ most central collisions are selected.
For this selection, the fluctuations in the number of target participants is also minimized~\cite{Konchakovski:2005hq}. Remarks on the contribution
of target participant fluctuations to multiplicity fluctuations are presented in~\cite{Gazdzicki:2005rr}.

A small fraction of the produced particles in a collision is also entering the calorimeter and introducing a small bias on centrality measurement.
Acceptance tables of the veto calorimeter as a function of $p$, $p_T$ and $\phi$ can be obtained on the author`s 
website (http://www.ikf.physik.uni-frankfurt.de/users/lungwitz/acceptance/) and are used for the UrQMD predictions of multiplicity fluctuations.

The predictions for the energy dependence of scaled variance measured in the NA49 experiment are shown in figures~\ref{w_hp}-\ref{w_hpm}.
To study the influence of centrality selection the scaled variance was also calculated for \Pb collisions with a zero impact parameter $b$.

The UrQMD model predicts a weak energy dependence of scaled variance for positively and negatively charged hadrons in forward acceptance.
At midrapidity, at full experimental acceptance and for all charged hadrons at all acceptances an increase of scaled variance with collision
energy is predicted.

In forward acceptance a UrQMD simulation for events with zero impact parameter ($b=0$) gives similar results to the simulation
for events selected according to their energy in the veto calorimeter. 
In midrapidity and in full experimental acceptance the scaled variance for events selected by their veto energy is larger, 
probably due to target participant fluctuations.

\section{Summary}

We present predictions for event-by-event multiplicity fluctuations for very central Pb+Pb and nucleon+nucleon 
interactions from $E_{lab}=20A$~GeV to $E_{lab}=158A$~GeV within a hadron-string transport approach. 
We find that the fluctuations do generally increase strongly towards higher beam energies, both for elementary and massive 
nuclear reactions. 
This can be used to distinguish the present model from a hadron gas model, 
which predicts a rather weak dependence of the fluctuations on energy. The amount of fluctuations in full phase 
space is generally non-Poissonian  (smaller at low energies, higher at high energies), crossing the Poissonian 
value in the SPS energy regime. 
Applying a forward rapidity cut, yields a non-monotonous behaviour as a function of energy, with a local minimum
at SPS energies. We further predict the rapidity dependence of the scaled variance at the highest SPS energy
and find a strong rapidity dependence of the fluctuations, even if trivial multiplicity effects are scaled out.
This might render the procedure to simply scale thermal model predictions for fluctuations to
the experimentally covered phase space questionable. The transverse momentum dependence of the fluctuations
does generally tend to decrease towards higher transverse momenta. This effect is related to energy conservation,
allowing stronger fluctuations for low energetic particles, while constraining the high energetic particles.
Finally, we analyse the influence of the veto trigger used by the NA49 experiment, 
compared to simple zero impact parameter reactions. Here we observe
a systematic deviation for the midrapidity results on the order of 10\%, at forward rapidities, the veto trigger
can be well approximated with a the zero impact parameter interaction. 

The present study therefore provides a detailed baseline calculation for the search of critical phenomena in event-by-event
multiplicity fluctuations. If non-monotonous deviations from these predictions, as e.g. expected by droplet
formation, are
observed, these enhanced fluctuations might indicate the onset of deconfinement and/or the critical point.

The NA49 collaboration is currently studying the energy dependence of multiplicity fluctuations from \enI to \eh. 
Further detailed exploration is planned for the NA61 (SHINE) experiment~\cite{Antoniou:2006mh} at the CERN SPS.
In addition the critRHIC experiment and the CBM experiment~\cite{Senger:2004jw} at the GSI FAIR facility as well as 
and the MPD experiment at JINR in Dubna will be 
able to explore this energy region in the near future.

\begin{acknowledgments}

The authors would like to thank M. Gazdzicki, M. Hauer, E. Bratkovskaya, V. Konchakovski, V. Begun, M. Gorenstein and I. Mishustin for 
fruitful discussions.
This work was partially supported by the Virtual Institute of Strongly Interacting Matter (VI-VH-146) of the Helmholtz-Gemeinschaft, the 
Bundesministerium f\"ur Bildung und Forschung (BMBF) and the Gesellschaft f\"ur Schwerionenforschung (GSI).

\end{acknowledgments}

\end{document}